\documentclass[twocolumn,aps,pra,amsmath,amssymb,notitlepage]{revtex4-1}
\usepackage[latin9]{inputenc}
\usepackage{mathrsfs}
\usepackage{amsmath}
\usepackage{amssymb}
\usepackage{graphicx}
\usepackage{esint}

\makeatletter

\newcommand*\LyXThinSpace{\,\hspace{0pt}}
\providecommand{\tabularnewline}{\\}

\usepackage{epsfig}
\usepackage{bm}
\usepackage{dcolumn}
\usepackage{color}

\makeatother

\begin{document}
\title{Microscopic pairing theory of a binary Bose mixture with interspecies
attractions: bosonic BEC-BCS crossover and ultradilute low-dimensional
quantum droplets}
\author{Hui Hu, Jia Wang, and Xia-Ji Liu}
\affiliation{Centre for Quantum Technology Theory, Swinburne University of Technology,
Melbourne, Victoria 3122, Australia}
\date{\today}
\begin{abstract}
Ultradilute quantum droplets are intriguing new state of matter, in
which the attractive mean-field force can be balanced by the repulsive
force from quantum fluctuations to avoid collapse. Here, we present
a microscopic theory of ultradilute quantum droplets in three-, one-
and two-dimensional two-component Bose-Bose mixtures, by generalizing
the conventional Bogoliubov theory to include the bosonic pairing
arising from the interspecies attraction. Our pairing theory is fully
equivalent to a variational approach and hence gives an upper bound
for the energy of quantum droplets. In three dimensions, we predict
the existence of a strongly interacting Bose droplet at the crossover
from Bose-Einstein condensates (BEC) to Bardeen--Cooper--Schrieffer
(BCS) superfluids and map out the bosonic BEC-BCS crossover phase
diagram. In one dimension, we find that the energy of the one-dimensional
Bose droplet calculated by the pairing theory is in an excellent agreement
with the latest diffusion Monte Carlo simulation {[}Phys. Rev. Lett.
\textbf{122}, 105302 (2019){]}, for nearly all the interaction strengths
at which quantum droplets exist. In two dimensions, we show that Bose
droplets disappear and may turn into a soliton-like many-body bound
state, when the interspecies attraction exceeds a critical value.
Below the threshold, the pairing theory predicts more or less the
same results as the Bogoliubov theory derived by Petrov and Astrakharchik
{[}Phys. Rev. Lett. \textbf{117}, 100401 (2016){]}. The predicted
energies from both theories are higher than the diffusion Monte Carlo
results, due to the weak interspecies attraction and the increasingly
important role played by the beyond-Bogoliubov-approximation effect
in two dimensions. Our pairing theory provides an ideal starting point
to understand interesting ground-state properties of quantum droplets
in various dimensions, including their shape and collective oscillations. 
\end{abstract}
\maketitle

\section{Introduction}

In the weakly interacting regime, quantum phase of ultracold atomic
Bose gases is typically determined by their mean-field interactions
\citep{Dalfovo1999}. Attractive mean-field interactions can induce
mechanical instability towards collapse \citep{Donley2001}. This
common viewpoint, however, is radically changed due to the seminal
work by Petrov \citep{Petrov2015}, who proposed that the mean-field
collapse could be prevented by the repulsive force provided by quantum
fluctuations, i.e., the celebrated Lee-Huang-Yang (LHY) correction
to the energy functional \citep{LeeHuangYang1957}. Although the beyond-mean-field
LHY correction is usually small, it can be made comparable to the
mean-field energy by experimentally tuning the interatomic interactions
with the Feshbach resonance technique. As a result, self-bound liquid-like
quantum droplets may form, even in free space without container \citep{Petrov2018,FerrierBarbut2019,Kartashov2019}.
Petrov's ground-breaking proposal has now been surprisingly confirmed
in single-component Bose gases with anisotropic dipolar forces \citep{FerrierBarbut2016,Schmitt2016,Chomaz2016,Bottcher2019PRA,Bottcher2019PRX}
and in two-component Bose-Bose mixtures with contact inter-particle
interactions \citep{Cabrera2018,Cheiney2018,Semeghini2018,Ferioli2019,DErrico2019}.
It opens a new rapidly developing research field \citep{Bottcher2020},
where the beyond-mean-field many-body effect could be systematically
explored, both experimentally \citep{FerrierBarbut2016,Schmitt2016,Chomaz2016,Bottcher2019PRA,Bottcher2019PRX,Cabrera2018,Cheiney2018,Semeghini2018,Ferioli2019,DErrico2019}
and theoretically \citep{Petrov2016,Baillie2016,Wachtler2016,Baillie2017,Li2017,Cappellaro2017,Jorgensen2018,Cui2018,Staudinger2018,Parisi2019,Cikojevic2019,Tylutki2020,Shi2019,Cikojevic2020,Wang2020,Hu2020a,Ota2020}.

Despite the great success of Petrov's proposal, strictly speaking,
it is not a consistent microscopic theory. This is particularly clear
for three-dimensional Bose-Bose mixtures, with which the Petrov prototype
theory of quantum droplets was constructed, within the Bogoliubov
approximation \citep{Petrov2015,Larsen1963}. As the mean-field solution
is not stable towards collapse, one of the two gapless Bogoliubov
modes becomes softened and acquires a small imaginary component. This
results in a complex LHY energy functional \citep{Petrov2015,Cikojevic2019,Hu2020a,Ota2020},
which is not physical. To circumvent this technical issue, Petrov
assumed a weak dependence of the LHY energy functional on the interspecies
interaction strength and fixed the LHY energy to the value on the
verge of the collapse where the mechanical instability first sets
in \citep{Petrov2015}. Hereafter, we will refer to such an approximation
as Petrov's prescription. Due to the intrinsic inconsistency in the
Petrov prototype theory, the resulting energy of quantum droplets
shows an appreciable deviation from the numerically accurate diffusion
Monte Carlo (DMC) predictions \citep{Cikojevic2019}. The predicted
critical number for the droplet formation also seems to be larger
than the one measured experimentally, both in dipolar Bose gases \citep{Bottcher2019PRA}
and in Bose-Bose mixtures \citep{Cabrera2018}.

Recently, we developed a \emph{consistent} microscopic theory to remove
the annoying loophole in the Petrov theory of quantum droplets \citep{Hu2020a}.
The crucial ingredient of the theory is the inclusion of a weak bosonic
pairing between different components due to the attractive interspecies
interactions. The pairing explicitly removes the unstable softened
Bogoliubov excitation and turns it into a stable gapped mode, as in
the conventional Bardeen-Cooper-Schrieffer (BCS) theory for interacting
fermions \citep{BCS1957}. An apparent advantage of the pairing theory
is that, it is variational and therefore predicts an \emph{upper}
bound for the ground-state energy. Remarkably, in three dimensions
our pairing theory leads to an improved agreement with the DMC simulation,
for the energy and the equilibrium density of quantum droplets \citep{Hu2020a}.

In this work, we would like to provide more details of the microscopic
pairing theory in three dimensions \citep{Hu2020a}, on the equation
derivation and the numerical calculation. In particular, we show that
the weakly interacting pairing in our previous study can be naturally
generalized to the strongly interacting regime, in which the interspecies
scattering length diverges across a Feshbach resonance. Therefore,
we predict the existence of a \emph{strongly interacting} Bose droplet,
where a significant fraction of atoms turns into bound pairs. This
is precisely analogous to the crossover physics from a Bose-Einstein
condensate (BEC) to a BCS superfluid \citep{NSR1985,Hu2006,Diener2008},
recently observed with fermionic $^{40}$K and $^{6}$Li atoms \citep{Regal2004,Zwierlein2004}.
Across the \emph{bosonic} BEC-BCS crossover, the interspecies scattering
length becomes positive and small, and the strongly interacting Bose
droplet eventually disappears and turns into a gas of tightly-bound
molecules or dimers via a first-order phase transition.

We then focus on the new cases of low-dimensional quantum droplets
and examine systematically their bulk properties. Our pairing theory
turns out to work extremely well in one dimension. For nearly all
the interaction strengths at which quantum droplets exist, we find
an excellent agreement between the theory and the DMC simulation for
the ground-state energy \citep{Parisi2019}. In two dimensions, quantum
droplets emerges for an arbitrarily weak interspecies interaction
strength. Due to the weakness of the interspecies attraction, our
pairing theory does not differ too much with the Petrov theory \citep{Petrov2016}.
Both theories fail to have a good agreement with the DMC simulation,
presumably due to the beyond-LHY-correction that becomes increasingly
important in two dimensions \citep{Mora2009,Astrakharchik2009}. Nevertheless,
it is remarkable that with increasing interspecies attraction our
pairing theory predict a critical attraction, above which quantum
droplets cease to exist. This critical value is consistent with the
threshold for zero-crossing in dimer-dimer scatterings from a four-body
problem in two dimensions \citep{Guijarro2020}. Above the threshold,
the effective interaction between dimers (i.e., tightly bound bosonic
pairs) changes from weakly attractive to weakly repulsive, indicating
the instability of quantum droplets in the few-body limit.

The rest of the paper is organized as follows. In the next section
(Sec. II), we introduce the model Hamiltonian for two-component Bose-Bose
mixtures and present the pairing theory, i.e, the Bogoliubov theory
with bosonic pairing. We mention briefly the concept of a bosonic
BEC-BCS crossover. In Sec. III, the connection of our pairing theory
to the conventional Bogoliubov theory without pairing is discussed.
In Sec. IV, Sec. V, and Sec. VI, we consider the three-, one-, and
two-dimensional cases, respectively. In three dimensions, we show
the existence of a strongly interacting Bose droplet and map out the
phase diagram of the bosonic BEC-BCS crossover. In the low-dimensional
cases, we discuss in detail the comparison of the pairing theory to
the benchmark DMC simulations and the physics near the threshold,
at which the Bose droplet starts to disappear. Finally, Sec. VII is
devoted to conclusions and outlooks.

\section{The model Hamiltonian and pairing theory}

We consider a two-component Bose-Bose mixture as in the seminal work
by Petrov \citep{Petrov2015}. To be specific, let us focus on homonuclear
mixtures such as the $^{39}$K-$^{39}$K mixture, with which the masses
of the two components are the same, i.e., $m_{1}=m_{2}=m$. In the
presence of the intraspecies interactions $g_{11}$ and $g_{22}$,
and the interspecies interactions $g_{12}=g_{21}$, the system in
free space can be described the following Hamiltonian density, 
\begin{eqnarray}
\mathscr{H}\left(\mathbf{x}\right) & = & \mathscr{H}_{0}\left(\mathbf{x}\right)+\mathscr{H}_{\textrm{int}}\left(\mathbf{x}\right),\\
\mathscr{H}_{0}\left(\mathbf{x}\right) & = & \sum_{i=1,2}\phi_{i}^{\dagger}\left(\mathbf{x}\right)\left[-\frac{\hbar^{2}\nabla^{2}}{2m}-\mu_{i}\right]\phi_{i}\left(\mathbf{x}\right),\\
\mathscr{H}_{\textrm{int}}\left(\mathbf{x}\right) & = & \sum_{i,j=1,2}\frac{g_{ij}}{2}\phi_{i}^{\dagger}\left(\mathbf{x}\right)\phi_{j}^{\dagger}\left(\mathbf{x}\right)\phi_{j}\left(\mathbf{x}\right)\phi_{i}\left(\mathbf{x}\right),
\end{eqnarray}
where $\phi_{i}(\mathbf{x})$ ($i=1,2$) is the annihilation field
operator of the $i$-species bosons and $\mu_{i}$ is the chemical
potential. In three or two dimensions, the use of contact inter-particle
interactions leads to the well-known ultraviolet divergence, so the
bare interaction strengths $g_{ij}$ need regularization and are to
be replaced by the $s$-wave scattering lengths $a_{ij}$ or the binding
energies $E_{B}^{ij}$. For example, in three dimensions we may write,
\begin{equation}
\frac{1}{g_{ij}}=\frac{m}{4\pi\hbar^{2}a_{ij}}-\frac{1}{\mathcal{V}}\sum_{k}\frac{m}{\hbar^{2}\mathbf{k}^{2}},\label{eq:Reg3d}
\end{equation}
where the volume $\mathcal{V}$ (or area in two dimensions and length
in one dimension) will be set to unity hereafter.

We use the conventional path-integral formalism to describe our bosonic
pairing theory, following its fermionic counterpart \citep{He2015,Hu2020b}.
We are interested in calculating the thermodynamic potential $\varOmega$
from the partition function, 
\begin{equation}
\mathcal{Z}=\int\mathcal{D}[\phi_{1},\phi_{2}]e^{-\mathcal{S}},
\end{equation}
where the action is given by, 
\begin{equation}
\mathcal{S}=\int dx\left[\sum_{i=1,2}\bar{\phi}_{i}\left(x\right)\partial_{\tau}\phi_{i}\left(x\right)+\mathscr{H}\left(x\right)\right].
\end{equation}
Here, we have used the standard notations $x\equiv(\mathbf{x},\tau)$
and $\int dx\equiv\int d\mathbf{x}\int_{0}^{\beta}d\tau$, and $\beta\equiv1/(k_{B}T)$.

Due to the attractive interspecies interaction (i.e, $g_{12}<0$),
we anticipate the pairing between different species. To make it evident,
we explicitly introduce an auxiliary pairing field $\Delta(x)$ and
take the Hubbard--Stratonovich transformation to decouple the Hamiltonian
density for interspecies interactions, i.e., 
\begin{widetext}
\begin{equation}
\exp\left[-g_{12}\int dx\bar{\phi}_{1}\bar{\phi}_{2}\phi_{2}\phi_{1}\right]=\int\mathcal{D}\left[\Delta\left(x\right)\right]\exp\left\{ \int dx\left[\frac{\left|\Delta\left(x\right)\right|^{2}}{g_{12}}+\left(\bar{\Delta}\phi_{2}\phi_{1}+\bar{\phi}_{1}\bar{\phi}_{2}\Delta\right)\right]\right\} .
\end{equation}
The action then becomes, 
\begin{equation}
\mathcal{S}=\int dx\left[-\frac{\left|\Delta\left(x\right)\right|^{2}}{g_{12}}-\left(\bar{\Delta}\phi_{2}\phi_{1}+\bar{\phi}_{1}\bar{\phi}_{2}\Delta\right)\right]+\sum_{i=1,2}\int dx\left[\bar{\phi}_{i}\left(\partial_{\tau}-\frac{\hbar^{2}\nabla^{2}}{2m}-\mu_{i}\right)\phi_{i}+\frac{g_{ii}}{2}\bar{\phi}_{i}^{2}\phi_{i}^{2}\right].
\end{equation}
\end{widetext}

For the pairing field $\Delta(x)$, it suffices to take a \emph{uniform}
saddle-point solution $\Delta(x)=\Delta>0$. At the same level of
the Bogoliubov approximation, at zero temperature we assume the bosons
condensate into the zero-momentum states, i.e., 
\begin{eqnarray}
\phi_{i}\left(x\right) & = & \phi_{ic}+\delta\phi_{i}\left(x\right),
\end{eqnarray}
with a positive wave-function $\phi_{ic}>0$. Following the Bogoliubov
approximation, the intraspecies interaction terms may be approximated
as, 
\begin{equation}
\frac{g_{ii}}{2}\bar{\phi}_{i}^{2}\phi_{i}^{2}\simeq\frac{C_{i}^{2}}{2g_{ii}}+\frac{C_{i}}{2}\left(4\delta\bar{\phi}_{i}\delta\phi_{i}+\delta\bar{\phi}_{i}\delta\bar{\phi}_{i}+\delta\phi_{i}\delta\phi_{i}\right),
\end{equation}
where $C_{i}=g_{ii}\phi_{ic}^{2}$. As a consequence, we find the
effective action $\mathcal{S}\simeq\beta\mathcal{V}\varOmega_{0}+\mathcal{S}_{B}$,
where the condensate thermodynamic potential $\varOmega_{0}$ is given
by, 
\begin{equation}
\varOmega_{0}=\sum_{i=1,2}\left(-\mu_{i}\phi_{ic}^{2}+\frac{C_{i}^{2}}{2g_{ii}}\right)-\frac{\Delta^{2}}{g_{12}}-2\Delta\phi_{1c}\phi_{2c},\label{eq:Omega0}
\end{equation}
and the quantum fluctuations around the condensates have the contribution,
\begin{align}
\mathcal{S}_{B} & =-\int dx\Delta\left(\delta\bar{\phi}_{1}\delta\bar{\phi}_{2}+\delta\phi_{2}\delta\phi_{1}\right)+\int dx\sum_{i=1,2}\nonumber \\
 & \left[\delta\bar{\phi}_{i}\left(\partial_{\tau}+\hat{B}_{i}\right)\delta\phi_{i}+\frac{C_{i}}{2}\left(\delta\bar{\phi}_{i}\delta\bar{\phi}_{i}+\delta\phi_{i}\delta\phi_{i}\right)\right],
\end{align}
with $\hat{B}_{i}(x)\equiv-\hbar^{2}\nabla^{2}/(2m)-\mu_{i}+2C_{i}$.
By introducing a Nambu spinor $\Phi(x)=[\delta\phi_{1}(x),\delta\bar{\phi}_{1}(x),\delta\phi_{2}(x),\delta\bar{\phi}_{2}(x)]^{T}$,
we may recast $\mathcal{S}_{B}$ into a compact form, 
\begin{equation}
\mathcal{S}_{B}=\int dxdx'\bar{\Phi}\left(x\right)\left[-\mathscr{D}^{-1}\left(x,x'\right)\right]\Phi\left(x'\right),
\end{equation}
where the inverse Green function of bosons is given by, 
\begin{equation}
\mathscr{D}^{-1}=\left[\begin{array}{cccc}
-\partial_{\tau}-\hat{B}_{1} & -C_{1} & 0 & \Delta\\
-C_{1} & \partial_{\tau}-\hat{B}_{1} & \Delta & 0\\
0 & \Delta & -\partial_{\tau}-\hat{B}_{2} & -C_{2}\\
\Delta & 0 & -C_{2} & \partial_{\tau}-\hat{B}_{2}
\end{array}\right].
\end{equation}
Due to the delta function $\delta\left(x-x'\right)$ in $\mathscr{D}^{-1}(x,x')$,
which we do not explicitly show in the above equation, it is convenient
to work in momentum space by performing a Fourier transform. After
replacing $-\partial_{\tau}$ with the bosonic Matasubara frequencies
$i\omega_{m}$ (i.e., $\omega_{m}=2\pi mk_{B}T$ with $m\subseteq\mathbb{Z}$)
and performing the analytic continuation $i\omega_{m}\rightarrow\omega+i0^{+}$,
i.e., 
\begin{equation}
-\partial_{\tau}\rightarrow\omega+i0^{+},
\end{equation}
and taking the replacement 
\begin{equation}
\hat{B}_{i}\rightarrow B_{i\mathbf{k}}=\varepsilon_{\mathbf{k}}-\mu_{i}+2C_{i}
\end{equation}
with $\varepsilon_{\mathbf{k}}=\hbar^{2}\mathbf{k}^{2}/(2m)$, it
is straightforward to explicitly write down the expression of $\mathscr{D}^{-1}(\mathbf{k},\omega)$.
By solving the poles of the bosonic Green function, i.e., $\det[\mathscr{D}^{-1}(\mathbf{k},\omega\rightarrow E(\mathbf{k}))]=0$,
or more explicitly,
\begin{widetext}
\begin{equation}
\omega^{4}-\omega^{2}\left[\left(B_{1\mathbf{k}}^{2}-C_{1}^{2}\right)+\left(B_{2\mathbf{k}}^{2}-C_{2}^{2}\right)-2\Delta^{2}\right]+\left[\left(B_{1\mathbf{k}}^{2}-C_{1}^{2}\right)\left(B_{2\mathbf{k}}^{2}-C_{2}^{2}\right)-2\left(B_{1\mathbf{k}}B_{2\mathbf{k}}+C_{1}C_{2}\right)\Delta^{2}+\Delta^{4}\right]=0,\label{eq:Det}
\end{equation}
we obtain the two Bogoliubov spectra, 
\begin{equation}
E_{\pm}^{2}\left(\mathbf{k}\right)=\left[\mathcal{A}_{+}\left(\mathbf{k}\right)-\Delta^{2}\right]\pm\sqrt{\mathcal{A}_{-}^{2}\left(\mathbf{k}\right)+\Delta^{2}\left[\left(C_{1}+C_{2}\right)^{2}-\left(B_{1\mathbf{k}}-B_{2\mathbf{k}}\right)^{2}\right]},
\end{equation}
\end{widetext}

where we have defined, 
\begin{equation}
\mathcal{A}_{\pm}\left(\mathbf{k}\right)=\frac{\left(B_{1\mathbf{k}}^{2}-C_{1}^{2}\right)\pm\left(B_{2\mathbf{k}}^{2}-C_{2}^{2}\right)}{2}.
\end{equation}

\subsection{Thermodynamic potential}

By taking the derivative of the condensate thermodynamic potential
$\varOmega_{0}$ in Eq. (\ref{eq:Omega0}) with respect to $\phi_{1c}$
and $\phi_{2c}$, we find that, 
\begin{eqnarray}
-\mu_{1}\phi_{1c}+g_{11}\phi_{1c}^{3}-\Delta\phi_{2c} & = & 0,\\
-\mu_{2}\phi_{2c}+g_{22}\phi_{2c}^{3}-\Delta\phi_{1c} & = & 0.
\end{eqnarray}
Therefore, we obtain 
\begin{eqnarray}
-\mu_{1}+C_{1}=B_{1\mathbf{k}=0}-C_{1} & = & \Delta\left(\phi_{2c}/\phi_{1c}\right),\\
-\mu_{2}+C_{2}=B_{2\mathbf{k}=0}-C_{2} & = & \Delta\left(\phi_{1c}/\phi_{2c}\right),
\end{eqnarray}
and hence 
\begin{equation}
\left(B_{1\mathbf{k}=0}-C_{1}\right)\left(B_{2\mathbf{k}=0}-C_{2}\right)=\Delta^{2}.
\end{equation}
As the last term in Eq. (\ref{eq:Det}) can be rewritten as the product
of $(B_{1\mathbf{k}}-C_{1})(B_{2\mathbf{k}}-C_{2})-\Delta^{2}$ and
$(B_{1\mathbf{k}}+C_{1})(B_{2\mathbf{k}}+C_{2})-\Delta^{2}$, the
term is zero at zero momentum $\mathbf{k}=0$. Thus, we confirm that
at least one of the two Bogoliubov spectra is gapless. This is anticipated
from the $U(1)$ symmetry breaking of the system. On the other hand,
it is straightforward to rewrite the condensate thermodynamic potential
in the form, 
\begin{equation}
\varOmega_{0}=-\frac{\Delta^{2}}{g_{12}}-\frac{C_{1}^{2}}{2g_{11}}-\frac{C_{2}^{2}}{2g_{22}}.
\end{equation}
We now turn to consider the action for quantum fluctuations $\mathcal{S}_{B}$,
which gives the LHY contribution to the thermodynamic potential \citep{Salasnich2016,Hu2020c},
\begin{align}
\varOmega_{\textrm{LHY}} & =\frac{k_{B}T}{2}\sum_{\mathbf{q},i\omega_{m}}\ln\det\left[\mathscr{-D}^{-1}\left(\mathbf{q},i\omega_{m}\right)\right]e^{i\omega_{m}0^{+}},\\
 & =\frac{1}{2}\sum_{\mathbf{k}}\left[E_{+}\left(\mathbf{k}\right)+E_{-}\left(\mathbf{k}\right)-B_{1\mathbf{k}}-B_{2\mathbf{k}}\right].
\end{align}
In two and three dimensions, it is worth noting that both $\varOmega_{0}$
and $\varOmega_{\textrm{LHY}}$ have ultraviolet divergence. However,
these two divergences can cancel with each other exactly, once the
regularization of the bare interaction strengths $g_{ij}$ is applied.
This will be discussed in more details when we explicitly write down
the total thermodynamic potential in different dimensions.

\subsection{Equal intraspecies interactions}

For simplicity, from now on, let us concentrate on the case with equal
intraspecies interactions $g_{11}=g_{22}=g$. With symmetric intraspecies
interactions, it is natural to take the same population for bosons
in different species, i.e., $\phi_{1c}=\phi_{2c}$. Therefore, we
have $C_{1}=C_{2}=C=\mu+\Delta>0$ and $B_{1\mathbf{k}}=B_{2\mathbf{k}}=B_{\mathbf{k}}=\varepsilon_{\mathbf{k}}+C+\Delta$.
It is easy to find the two Bogoliubov spectra, 
\begin{eqnarray}
E_{-}(\mathbf{k}) & = & \sqrt{\varepsilon_{\mathbf{k}}\left(\varepsilon_{\mathbf{k}}+2C+2\Delta\right)},\label{eq:LowerEk}\\
E_{+}(\mathbf{k}) & = & \sqrt{\left(\varepsilon_{\mathbf{k}}+2C\right)\left(\varepsilon_{\mathbf{k}}+2\Delta\right)}.\label{eq:UpperEk}
\end{eqnarray}
The upper Bogoliubov branch $E_{+}(\mathbf{k})$ is thereby gapped,
provided the pairing gap $\Delta\neq0$. Finally, the total thermodynamic
potential takes the form, 
\begin{equation}
\varOmega=-\frac{C^{2}}{g}-\frac{\Delta^{2}}{g_{12}}+\frac{1}{2}\sum_{\mathbf{k}}\left[E_{+}\left(\mathbf{k}\right)+E_{-}\left(\mathbf{k}\right)-2B_{\mathbf{k}}\right].\label{eq:OmegaPairing}
\end{equation}
For a given chemical potential $\mu$, the saddle-point value of the
pairing gap $\Delta_{0}$ is to be determined by minimizing the thermodynamic
potential, i.e., 
\begin{equation}
\left.\frac{\partial\varOmega}{\partial\Delta}\right|_{\Delta_{0}}=0.
\end{equation}
We then calculate the total number of bosons in the droplet using
the number equation, 
\begin{equation}
n=-\frac{\partial\varOmega\left(\mu,\Delta_{0}\right)}{\partial\mu},
\end{equation}
and obtain the total energy of the droplet $E=\varOmega+n\mu$.

\subsection{Bosonic BEC-BCS crossover}

Our bosonic pairing theory for a binary Bose mixture runs in parallel
with the standard BCS theory for a two-component Fermi superfluid.
The only difference is that, there are positive intraspecies interactions
in the Bose mixture and each Bose component can individually undergo
Bose-Einstein condensation. The condensation leads to two immediate
consequences. First, the associated $U(1)$ symmetry breaking ensures
a gapless Bogoliubov spectrum $E_{-}(\mathbf{k})$, as we already
see from Eq. (\ref{eq:LowerEk}). On the other hand, the bosonic pairing
is mainly contributed from the zero momentum condensate state. As
a result, the pairing order parameter $\Delta$ becomes comparable
to the parameter $C$, which characterizes the typical interaction
energy scale. In other words, in the weak interacting limit (i.e.,
$\left|na_{12}^{3}\right|\ll1$) the pairing parameter is not exponentially
small as in the standard BCS theory. Yet, it is still small compared
with the ``Fermi'' energy, i.e., $\hbar^{2}n^{2/3}/(2m)$, associated
with the density $n$. Therefore, the gapped Bogoliubov spectrum $E_{+}(\mathbf{k})$
and the pairing gap $\Delta$ might be difficult to experimentally
probe by using the radio-frequency spectroscopy \citep{Chin2004}
and Bragg spectroscopy \citep{Stenger1999}.

The pairing gap could be enlarged significantly if we increase the
interspecies attraction $a_{12}$ by sweeping the magnetic field across
the Feshbach resonance \citep{Cabrera2018,Semeghini2018}. In two-component
Fermi gases of $^{40}$K and $^{6}$Li atoms, this leads to the so-called
BEC-BCS crossover \citep{NSR1985,Hu2006,Diener2008}, which has been
extensively studied over the past two decades \citep{Regal2004,Zwierlein2004,Chin2004}.
The same BEC-BCS crossover should occur, if the binary Bose mixture
is not destroyed by the three-body loss \citep{Cabrera2018,Semeghini2018}.
Approaching the Feshbach resonance, where the interspecies scattering
length $a_{12}$ diverges, we anticipate that the loosely-bound bosonic
Cooper pairs gradually shrink their size and become more tightly bound.
We then obtain a strongly interacting Bose droplet, with a significant
portion of pairs that behave like true molecules or dimers. At this
point, our mean-field theory becomes less accurate but is still qualitatively
reliable \citep{Hu2006,Diener2008}. Quantum fluctuations around the
mean-field pairing order parameter should be included for a quantitative
description. They give rise to another gapless branch in the collective
excitation spectrum associated with the $U$(1) symmetry breaking
of the pairing field, which has a sizable contribution to the thermodynamic
potential of the system \citep{Hu2006,Diener2008}. Across the Feshbach
resonance, $a_{12}$ becomes positive and starts to decrease. Eventually,
the strongly interacting Bose droplet ceases to exist and we should
instead obtain a gas of dimers. Numerical results of the bosonic BEC-BCS
crossover will be discussed in detail in Sec. IV.

\section{Bogoliubov theory and Petrov's prescription}

It is useful to explicitly compare the structure of our pairing theory
with that of the widely-used Petrov theory. For this purpose, here
we briefly review the Petrov's prototype theory of quantum droplets.
Within the Bogoliubov approximation \citep{Larsen1963}, we decouple
the interspecies interaction Hamiltonian density ($\phi_{1c}=\phi_{2c}=\phi_{c}$),
\begin{eqnarray}
g_{12}\bar{\phi}_{1}\bar{\phi}_{2}\phi_{2}\phi_{1} & \simeq & \frac{D^{2}}{g_{12}}-D\left(\delta\bar{\phi}_{1}\delta\phi_{1}+\delta\bar{\phi}_{2}\delta\phi_{2}\right)\nonumber \\
 &  & -D\left(\delta\bar{\phi}_{1}\delta\bar{\phi}_{2}+\delta\bar{\phi}_{1}\delta\phi_{2}+\textrm{H.c.}\right),
\end{eqnarray}
where H.c. stands for taking the Hermitian conjugate and $D\equiv-g_{12}\phi_{c}^{2}>0$
seems to play the role of the pairing field $\Delta$ in our pairing
theory. However, there is a slight difference. The Bogoliubov decoupling
shown in the above generates two \emph{additional} terms in the quantum
fluctuation action $\mathcal{S}_{B}$, i.e., $-\int dxD(\delta\bar{\phi}_{1}\delta\phi_{1}+\delta\bar{\phi}_{2}\delta\phi_{2})$
and $-\int dxD(\delta\bar{\phi}_{1}\delta\phi_{2}+\delta\bar{\phi}_{2}\delta\phi_{1})$.
In momentum space, the inverse Green function of bosons $\mathscr{D}^{-1}(\mathbf{k},\omega)$
then becomes, 
\begin{equation}
\mathscr{D}^{-1}=\left[\begin{array}{cccc}
\omega-B_{\mathbf{k}} & -C & D & D\\
-C & -\omega-B_{\mathbf{k}} & D & D\\
D & D & \omega-B_{\mathbf{k}} & -C\\
D & D & -C & -\omega-B_{\mathbf{k}}
\end{array}\right],
\end{equation}
where $B_{\mathbf{k}}\equiv\varepsilon_{k}-\mu+2C-D$. The existence
of the two additional terms leads to the two Bogoliubov spectra, 
\begin{equation}
\tilde{E}_{\pm}\left(\mathbf{k}\right)=\sqrt{\left(B_{\mathbf{k}}-C\right)\left(B_{\mathbf{k}}+C\mp2D\right)}.
\end{equation}
Here, we have used the \emph{tilde} to distinguish the dispersion
relations from those of the pairing theory. In this case, it is easy
to see that the condensate thermodynamic potential takes the form,
\begin{equation}
\varOmega_{0}=-2\mu\phi_{c}^{2}+g\phi_{c}^{4}+g_{12}\phi_{c}^{4}.
\end{equation}
By minimizing $\varOmega_{0}$ with respect to $\phi_{c}^{2}$, we
obtain the restriction, 
\begin{equation}
\mu=g\phi_{c}^{2}+g_{12}\phi_{c}^{2}=C-D,
\end{equation}
or equivalently $B_{\mathbf{k}=0}=C$, which ensures the gapless Bogoliubov
spectra. Therefore, we may rewrite down the condensate thermodynamic
potential, 
\begin{equation}
\varOmega_{0}=-\frac{C^{2}}{g}-\frac{D^{2}}{g_{12}},
\end{equation}
the two dispersion relations, 
\begin{eqnarray}
\tilde{E}_{-}(\mathbf{k}) & = & \sqrt{\varepsilon_{\mathbf{k}}\left(\varepsilon_{\mathbf{k}}+2C+2D\right)},\\
\tilde{E}_{+}(\mathbf{k}) & = & \sqrt{\varepsilon_{\mathbf{k}}\left(\varepsilon_{\mathbf{k}}+2C-2D\right)},
\end{eqnarray}
and also the LHY thermodynamic potential, 
\begin{equation}
\varOmega_{\textrm{LHY}}=\frac{1}{2}\sum_{\mathbf{k}}\left[\tilde{E}_{+}\left(\mathbf{k}\right)+\tilde{E}_{-}\left(\mathbf{k}\right)-2\left(\varepsilon_{\mathbf{k}}+C\right)\right].\label{eq:OmegaLHYBog}
\end{equation}
In three dimensions, using Eq. (\ref{eq:Reg3d}) we replace the bare
interaction strengths $g$ and $g_{12}$ with the $s$-wave scattering
lengths $a$ and $a_{12}$, respectively. Therefore, we obtain $\varOmega_{\textrm{3D}}=\varOmega_{0}+\varOmega_{\textrm{LHY}}$,
\begin{eqnarray}
\varOmega_{\textrm{3D}} & = & -\frac{m}{4\pi\hbar^{2}}\left[\frac{C^{2}}{a}+\frac{D^{2}}{a_{12}}\right]+\frac{1}{2}\sum_{\mathbf{k}}\left[\tilde{E}_{+}\left(\mathbf{k}\right)+\tilde{E}_{-}\left(\mathbf{k}\right)\right.\nonumber \\
 &  & \left.-2\left(\varepsilon_{\mathbf{k}}+C\right)+\frac{2\left(C^{2}+D^{2}\right)}{\hbar^{2}\mathbf{k}^{2}/m}\right].
\end{eqnarray}
The integration over the momentum can be easily calculated, by using
the identity 
\begin{equation}
\sum_{\mathbf{k}}\left[\sqrt{\varepsilon_{\mathbf{k}}\left(\varepsilon_{\mathbf{k}}+\alpha\right)}-\varepsilon_{\mathbf{k}}-\frac{\alpha}{2}+\frac{\alpha^{2}}{8\varepsilon_{\mathbf{k}}}\right]=\frac{\left(2m\right)^{3/2}\alpha^{5/2}}{15\pi^{2}\hbar^{3}}\label{eq:identity3d}
\end{equation}
in three dimensions. We arrive at, 
\begin{equation}
\varOmega_{\textrm{3D}}=-\frac{m}{4\pi\hbar^{2}}\left[\frac{C^{2}}{a}+\frac{D^{2}}{a_{12}}\right]+\frac{8m^{3/2}C^{5/2}}{15\pi^{2}\hbar^{3}}\mathcal{F}_{3}\left(\frac{D}{C}\right),\label{eq:OmegaPetrov}
\end{equation}
where $\mathcal{F}_{3}(\alpha)\equiv(1+\alpha)^{5/2}+(1-\alpha)^{5/2}$.
To calculate the total energy, we note that unlike the pairing gap
$\Delta$ in our pairing theory, the variable $D$ is not a variational
parameter. Therefore, there is an \emph{ambiguity} to determine the
variables $D$ and then $C=\mu+D$ for a given chemical potential
$\mu$. Nevertheless, we may assume that 
\begin{equation}
\frac{D}{C}=-\frac{g_{12}}{g}\simeq-\frac{a_{12}}{a},
\end{equation}
so that $C=\mu a/(a+a_{12})$ and $D=-\mu a_{12}/(a+a_{12})$. As
quantum droplets emerge when $D/C=-a_{12}/a>1$ in three dimensions
\citep{Petrov2015}, we immediately find that the Bogoliubov spectrum
$\tilde{E}_{+}(\mathbf{k})=\sqrt{\varepsilon_{\mathbf{k}}(\varepsilon_{\mathbf{k}}+2C-2D})$
becomes complex and consequently the function $\mathcal{F}_{3}(\alpha)$
in Eq. (\ref{eq:OmegaPetrov}) is ill-defined \citep{Petrov2015,Cikojevic2019,Hu2020a}.
To cure this problem, we may simply set $\alpha=D/C=1$ in the function
$\mathcal{F}_{3}(\alpha)$ \citep{Petrov2015} and hence the LHY term
become independent on the interspecies interaction strength. This
Petrov's prescription is now widely taken in the theoretical studies
of quantum droplets. We note also that, to calculate the total energy,
we may further approximate $C=(2\pi\hbar^{2}a/m)n$ and $\mu=[2\pi\hbar^{2}(a+a_{12})/m]n$,
which leads to the total energy, 
\begin{equation}
\frac{E_{\textrm{3D}}}{N}=\frac{\pi\hbar^{2}\left(a+a_{12}\right)}{m}n+\frac{256\sqrt{\pi}}{15}\frac{\hbar^{2}a^{5/2}}{m}n^{3/2}.\label{eq:EnergyPetrov3d}
\end{equation}
In other words, to determine the density $n$, we take the derivative
of the first term in Eq. (\ref{eq:OmegaPetrov}) only with respect
to the chemical potential $\mu$. This approximation is well justified
for a conventional weakly-interacting Bose gas in three dimensions
\citep{Salasnich2016}. However, it may not be convincing for quantum
droplets, where the second term in Eq. (\ref{eq:OmegaPetrov}) may
become comparable to the first term.

\section{Three-dimensional quantum droplets}

Let us now consider the pairing theory in three dimensions, showing
some technical details behind our previous work \citep{Hu2020a} and
describing the strongly interacting Bose droplet at the BEC-BCS crossover.
By regularizing the bare interaction strengths $g$ and $g_{12}$
in terms of the $s$-wave scattering lengths $a$ and $a_{12}$, we
rewrite the thermodynamic potential Eq. (\ref{eq:OmegaPairing}) into
the form, 
\begin{eqnarray}
\varOmega_{\textrm{3D}} & = & -\frac{m}{4\pi\hbar^{2}}\left[\frac{C^{2}}{a}+\frac{\Delta^{2}}{a_{12}}\right]+\frac{1}{2}\left(\mathcal{I}_{+}+\mathcal{I}_{-}\right),\\
\mathcal{I}_{\pm} & = & \sum_{\mathbf{k}}\left[E_{\pm}\left(\mathbf{k}\right)-\left(\varepsilon_{\mathbf{k}}+C+\Delta\right)+\frac{\left(C\pm D\right)^{2}}{2\varepsilon_{\mathbf{k}}}\right].
\end{eqnarray}
$\mathcal{I}_{-}$ can be directly calculated, with the help of the
identity Eq. (\ref{eq:identity3d}),

\begin{equation}
\mathcal{I}_{-}=\frac{16m^{3/2}}{15\pi^{2}\hbar^{3}}C^{5/2}\left(1+\frac{\Delta}{C}\right)^{5/2}.
\end{equation}
To calculate $\mathcal{I}_{+}$, we introduce a new variable $t\equiv[\hbar^{2}k^{2}/(2m)]/(2C)$
and $\alpha\equiv\Delta/C$ and rewrite

\begin{equation}
\mathcal{I}_{+}=\frac{16m^{3/2}}{15\pi^{2}\hbar^{3}}C^{5/2}h_{3}\left(\alpha\right),
\end{equation}
where the function 
\begin{eqnarray}
h_{3}\left(\alpha\right) & \equiv & \frac{15}{4}\int_{0}^{\infty}dt\sqrt{t}\left[\sqrt{\left(t+1\right)\left(t+\alpha\right)}-t\right.\nonumber \\
 &  & \left.-\frac{1+\alpha}{2}+\frac{\left(1-\alpha\right)^{2}}{8t}\right].
\end{eqnarray}
By combining $\mathcal{I}_{+}$ and $\mathcal{I}_{-}$, we obtain
the (regularized) LHY thermodynamic potential ($C=\mu+\Delta$), 
\begin{equation}
\varOmega_{\textrm{LHY}}=\frac{8m^{3/2}}{15\pi^{2}\hbar^{3}}\left(\mu+\Delta\right)^{5/2}\mathcal{G}_{3}\left(\frac{\Delta}{\mu+\Delta}\right),
\end{equation}
where $\mathcal{G}_{3}(\alpha)\equiv(1+\alpha)^{5/2}+h_{3}(\alpha)$.
Compared with the function $\mathcal{F}_{3}(\alpha)\equiv(1+\alpha)^{5/2}+(1-\alpha)^{5/2}$
in the last section, we find interestingly that the role of $(1-\alpha)^{5/2}$,
which is not well-defined for $\alpha>1$, is now taken by the new
function $h_{3}(\alpha)$. Therefore, we obtain the total thermodynamic
potential,

\begin{eqnarray}
\varOmega_{\textrm{3D}} & = & -\frac{m}{4\pi\hbar^{2}}\left[\frac{\left(\mu+\Delta\right)^{2}}{a}+\frac{\Delta^{2}}{a_{12}}\right]\nonumber \\
 &  & +\frac{8m^{3/2}}{15\pi^{2}\hbar^{3}}\left(\mu+\Delta\right)^{5/2}\mathcal{G}_{3}\left(\frac{\Delta}{\mu+\Delta}\right).\label{eq:OmegaPairing3d}
\end{eqnarray}
It takes essentially the same form as the thermodynamic potential
Eq. (\ref{eq:OmegaPetrov}) in the standard Bogoliubov theory, except
that the ill-defined function $\mathcal{F}_{3}(\alpha)$ is now replaced
by $\mathcal{G}_{3}(\alpha)$, and the pairing gap $\Delta$ is variational
and should be determined by minimizing $\varOmega_{\textrm{3D}}(\Delta)$.
For a given chemical potential $\mu$, we therefore minimize $\varOmega_{\textrm{3D}}$
to find the saddle-point value of the pairing order parameter $\Delta_{0}$.
For nonzero $\Delta_{0}\neq0$, we obtain $\varOmega_{\textrm{3D}}(\mu,\Delta_{0})$
and calculate $n=-\partial\varOmega_{\textrm{3D}}(\mu,\Delta_{0})/\partial\mu$.

In the weakly interacting regime, where $a+a_{12}\simeq0$, we typically
find that the chemical potential is much smaller than either the parameter
$C$ or the pairing gap $\Delta_{0}$ \citep{Hu2020a}. This could
be easily understood from the $\Delta$-dependence of $\varOmega_{0}$
and $\varOmega_{\textrm{LHY}}$, as shown in Eq. (\ref{eq:OmegaPairing3d}).
We note that two terms in $\varOmega_{0}$ are large and have opposite
sign. Each of them (i.e., absolute value) is much larger than $\varOmega_{\textrm{LHY}}$.
Therefore, when we minimize $\varOmega$ with respect to $\Delta$,
we only need to minimize $\Omega_{0}$. This leads to the condition,
\begin{equation}
\frac{\mu+\Delta_{0}}{a}+\frac{\Delta_{0}}{a_{12}}\simeq0.
\end{equation}
Hence, as a result of $a_{12}\sim-a$, we obtain 
\begin{equation}
\mu\simeq-\left(1+\frac{a}{a_{12}}\right)\Delta_{0}\ll\Delta_{0},C.
\end{equation}
Due to the smallness of $\left|\mu\right|$, it is reasonable to neglect
the $\mu$-dependence in $\varOmega_{\textrm{LHY}}$ and also the
term $\mu^{2}$ in $\varOmega_{0}$. Therefore, we obtain, 
\begin{equation}
\varOmega_{\textrm{3D}}\simeq-\frac{m}{4\pi\hbar^{2}}\left[\frac{2\mu\Delta+\Delta^{2}}{a}+\frac{\Delta^{2}}{a_{12}}\right]+\frac{32\sqrt{2}m^{3/2}}{15\pi^{2}\hbar^{3}}\Delta^{5/2}.
\end{equation}
By taking the derivative with respect to $\mu$ and taking the saddle-point
value $\Delta=\Delta_{0}$, we find 
\begin{equation}
n=-\frac{\partial\varOmega_{\textrm{3D}}}{\partial\mu}\simeq\frac{m}{2\pi\hbar^{2}a}\Delta_{0}.
\end{equation}
Replacing $\Delta_{0}$ by the density $n$ everywhere in $\varOmega_{\textrm{3D}}$
and calculate $E_{\textrm{3D}}=\varOmega_{\textrm{3D}}+\mu n$, we
finally arrive at an \emph{approximate} energy for small densities,
\begin{equation}
\frac{E_{\textrm{3D}}}{N}=-\frac{\pi\hbar^{2}}{m}\left(a+\frac{a^{2}}{a_{12}}\right)n+\frac{256\sqrt{\pi}}{15}\frac{\hbar^{2}a^{5/2}}{m}n^{3/2}.\label{eq:EnergyPairing3d}
\end{equation}
It is useful to compare this analytic expression with the energy functional
obtained by Petrov using his prescription \citep{Petrov2015}, i.e.,
Eq. (\ref{eq:EnergyPetrov3d}). It is interesting to see that these
two energy functionals have the exactly \emph{same} LHY term. The
reproduce of the Petrov's approximate LHY term in our pairing theory
suggests that Petrov's prescription is actually very reasonable, at
least for the case of equal intraspecies interactions considered here
\citep{Hu2020a}. However, it is worth emphasizing that, quite unexpectedly,
the mean-field energy term in our pairing theory (i.e., the first
term in Eq. (\ref{eq:EnergyPairing3d})) changes a lot. It is weakened
by a factor of $-a/a_{12}<1$, compared with the conventional mean-field
expression used by Petrov \citep{Petrov2015}, i.e., 
\begin{equation}
\left(g+g_{12}\right)\frac{n}{4}\rightarrow\frac{\pi\hbar^{2}(a+a_{12})}{m}n.
\end{equation}
This difference partly comes from our \emph{regularization} of the
bare interaction strengths, which is rigorously treated in the pairing
theory. As the beyond-mean-field LHY effect becomes dominant in quantum
droplets, a consistent treatment of the potential regularization is
necessary. Therefore, it is not a surprise why our regularized mean-field
energy becomes different from the widely-accepted conventional expression.

\begin{figure}[t]
\begin{centering}
\includegraphics[width=0.5\textwidth]{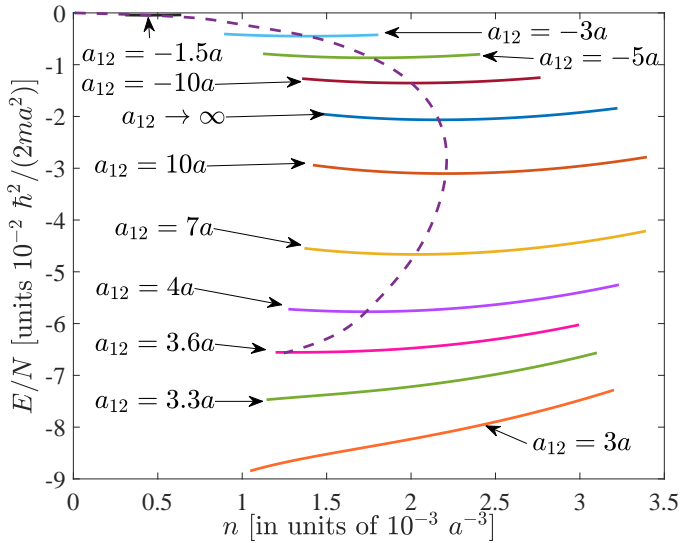} 
\par\end{centering}
\caption{\label{fig_crossover3d} Three-dimensional energy per particle $E/N$
of a strongly interacting Bose droplet predicted by the pairing theory,
as a function of the density $n$, with increasing interspecies attraction
from $a_{12}=-1.5a$ to $a_{12}=\mp\infty$, and finally to $a_{12}=+3.0a$,
as indicated in the figure. The dashed line traces the equilibrium
density. It ends at $a_{12}\simeq3.6a$, where the strongly interacting
Bose droplet disappears. The energy per particle $E/N$ is measured
in units of $10^{-2}\hbar^{2}/(2ma^{2})$ and the density $n$ is
in units of $10^{-3}a^{-3}$.}
\end{figure}

Let us now consider the strongly interacting regime, where interspecies
scattering length $\left|a_{12}\right|$ is significantly larger than
the intraspecies scattering length $a$. In Fig. \ref{fig_crossover3d},
we report the density dependence of the energy per particle $E/N$,
when the interspecies attraction increases from $a_{12}=-1.5a$ to
$a_{12}=-\infty$, crosses the Feshbach resonance (i.e., $a_{12}$
jumps from $-\infty$ to $+\infty$), and further increases towards
the formation of tightly bound molecules ($a_{12}\rightarrow0^{+}$).
Remarkably, with increasing interspecies attraction, the energy per
particle decreases steadily from the weakly interacting limit (i.e.,
$a_{12}<-1.5a$, see the top left of the figure, which has been studied
in Ref. \citep{Hu2020a}) to the strongly interacting regime ($a_{12}\rightarrow\pm\infty$).
This is very similar to what happens at the fermionic BEC-BCS crossover
\citep{NSR1985,Hu2006,Diener2008}, which is realized by tuning the
attractive interaction between two unlike fermions across a Feshbach
resonance \citep{Regal2004,Zwierlein2004}. Here, we always find a
minimum in the energy per particle (see the dashed line), until the
interspecies scattering length becomes positive and smaller than a
threshold $a_{12,\textrm{crit}}\simeq3.6a$. The existence of a minimum
in the energy per particle, i.e., $\partial(E/N)/\partial n=(\mu-E/N)/n=0$,
precisely implies the appearance of a self-bound Bose droplet in a
vacuum, which has a zero pressure $P=n\mu-E/\mathcal{V}=0$. Therefore,
we observe that a Bose droplet may exist in the strongly interacting
regime, where the interspecies scattering length diverges. Importantly,
the equilibrium density of the droplet, at which the minimum energy
per particle is reached, does not significantly increase as the interspecies
scattering length diverges. We find that the equilibrium density attains
its largest value near the Feshbach resonance and is about a few times
larger than the equilibrium density in the weakly coupling regime
(i.e., at $a_{12}=-1.5a$).

\begin{figure}[t]
\begin{centering}
\includegraphics[width=0.5\textwidth]{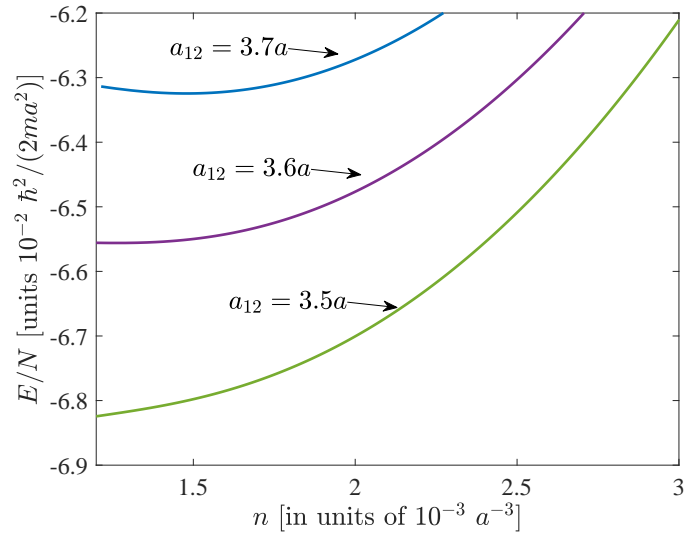} 
\par\end{centering}
\caption{\label{fig_critical3d} Three-dimensional energy per particle $E/N$
as a function of the density $n$, near the threshold $a_{12}=+3.6a$.}
\end{figure}

The experimental confirmation of a strongly interacting Bose droplet
would be a highly non-trivial task. In a single-component Bose gas,
it is well-known that a strongly interacting Bose gas is difficult
to reach due to severe three-body loss. Indeed, in the recent experiments
the lifetime of a $^{39}$K-$^{39}$K Bose droplet is limited to about
ten milliseconds, due to the \emph{intraspecies} three-body loss in
a particular component (i.e., the $\left|F=1,m_{F}=0\right\rangle $
hyperfine state) \citep{Cabrera2018,Semeghini2018}. This is not a
serious problem, since we can choose a binary Bose mixture with small
intraspecies three-body loss rate, for example, the heteronuclear
$^{41}$K-$^{87}$Rb mixture, where the lifetime of the droplet was
found to be much longer \citep{DErrico2019}. As the equilibrium density
of a strongly interacting Bose droplet is at the same order as a weakly
interacting droplet in magnitude, a large intraspecies three-body
loss could be avoided. The fundamental difficulty then may come from
the \emph{interspecies} three-body loss, which remains unknown at
the moment. Theoretically, the solution of the three-body problem
of two like bosons and one unlike boson and the calculation of the
related interspecies three-body loss rate would be an interesting
research topic to consider in future works.

\begin{figure}[t]
\begin{centering}
\includegraphics[width=0.5\textwidth]{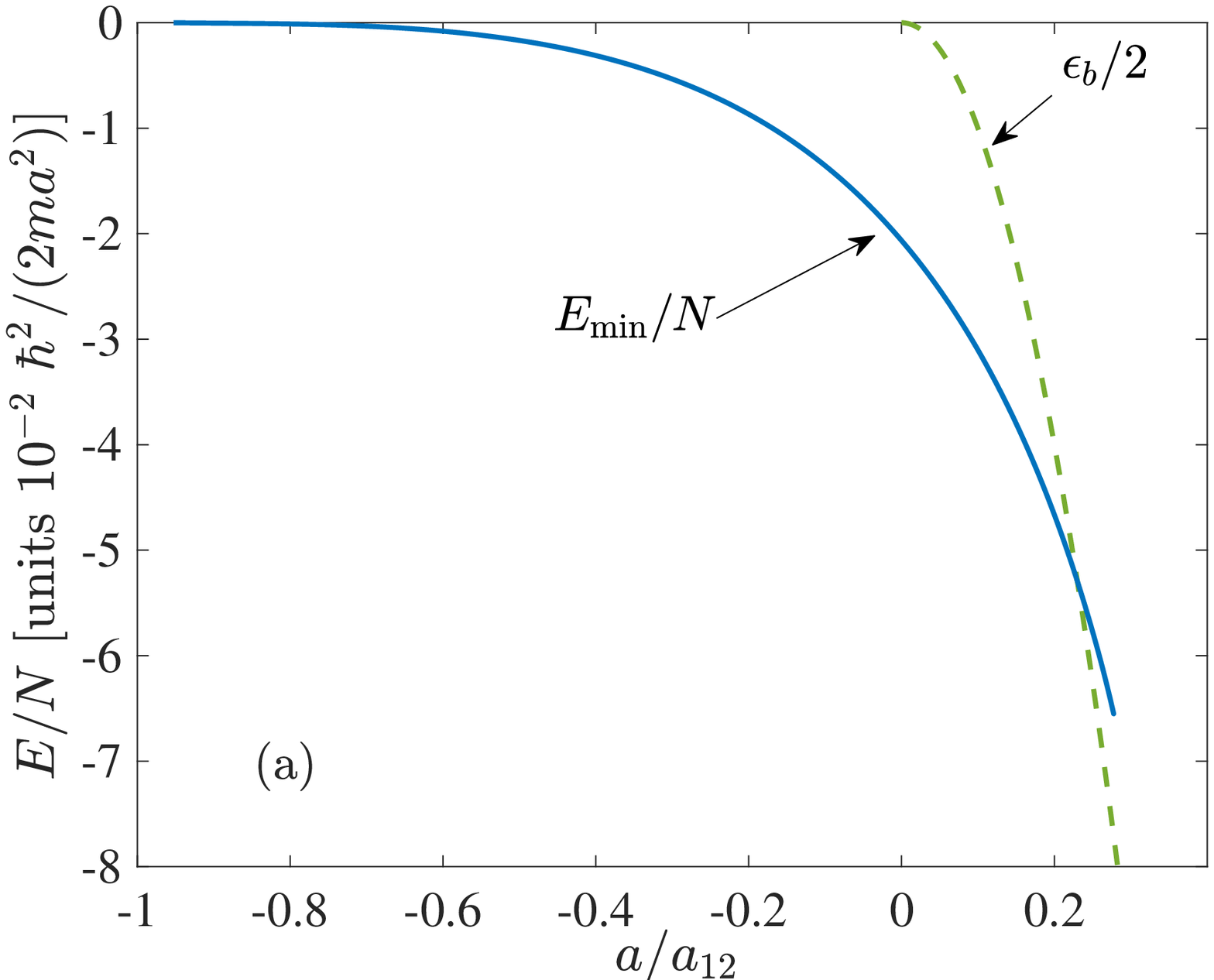} 
\par\end{centering}
\begin{centering}
\includegraphics[width=0.5\textwidth]{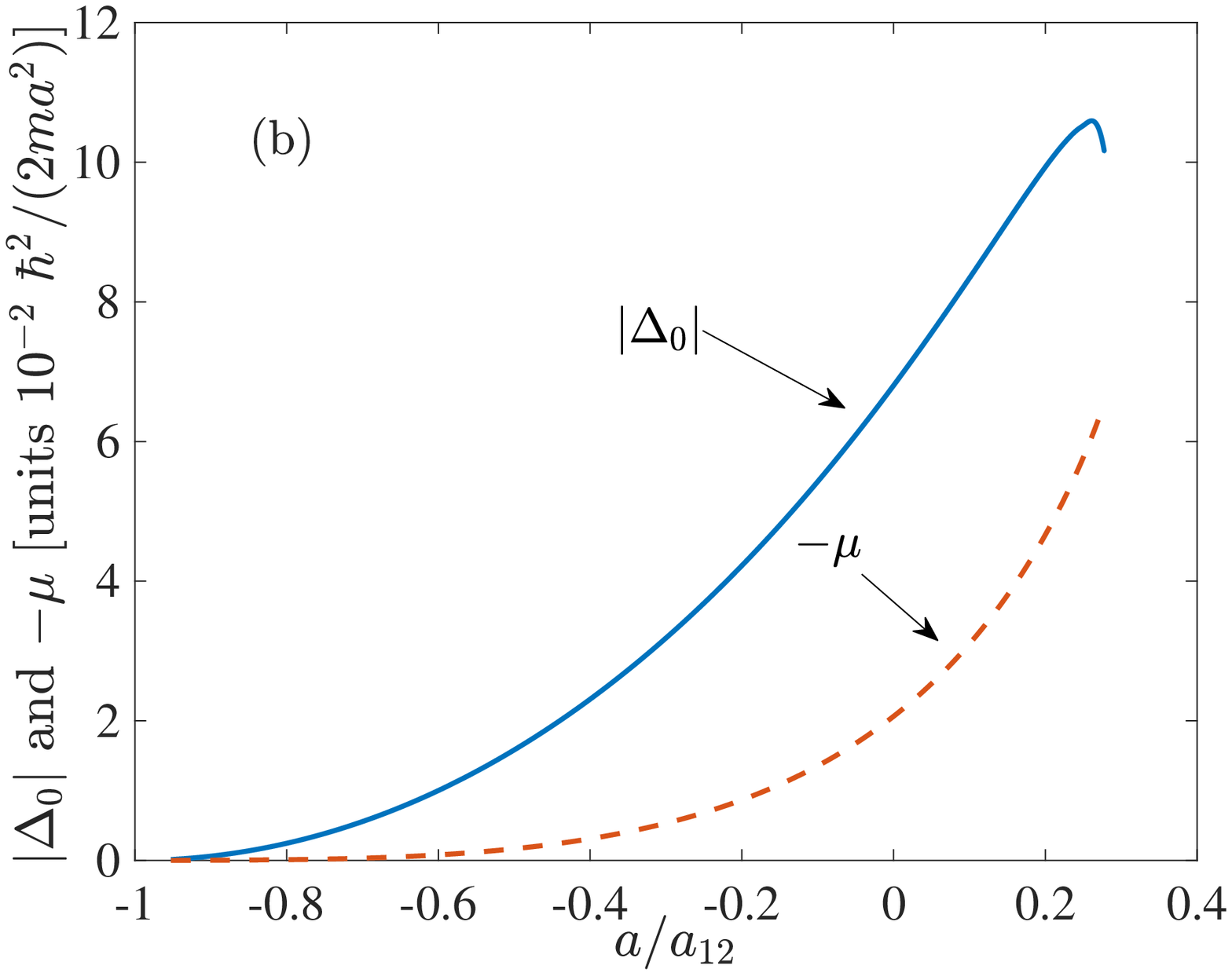} 
\par\end{centering}
\caption{\label{fig3_parameters3d} (a) Three-dimensional minimum energy per
particle $E_{\min}/N$ as a function of the ratio $a/a_{12}$. The
green dashed line shows the half of the bound-state energy $\epsilon_{B}/2$
of a molecule or dimer, formed by two bosons at different species.
(b) The pairing order parameter $\Delta_{0}$ (solid line) and the
chemical potential $-\mu$ (dashed line), as a function of the ratio
$a/a_{12}$. The minimum energy per particle $E_{\min}/N$, the pairing
order parameter $\Delta_{0}$, and the chemical potential $-\mu$
are all measured in units of $10^{-2}\hbar^{2}/(2ma^{2})$.}
\end{figure}

On the BEC side of the Feshbach resonance, the Bose droplet ceases
to exist below the threshold $a_{12,\textrm{crit}}\simeq3.6a$, as
highlighted in Fig. \ref{fig_critical3d}. This is closely related
to the formation of a gas of dimers towards the BEC limit, which also
happens at the fermionic BEC-BCS crossover \citep{NSR1985,Hu2006,Diener2008}.
To show this, in Fig. \ref{fig3_parameters3d}(a) we present the minimum
energy per particle $E_{\min}/N$ and the half of the bound-state
energy $\epsilon_{B}/2$ of a dimer, as a function of the ratio $a/a_{12}$.
It is readily seen that a gas of non-interacting dimers becomes energetically
favorable over a Bose droplet once we pass the crossing point of the
two energy curves at $a/a_{12}\simeq0.23$ or $a_{12}\simeq4.35a$.
Interestingly, the pairing order parameter $\Delta_{0}$ appears to
have a maximum around the crossing point, as shown in Fig. \ref{fig3_parameters3d}(b).

\begin{figure}[t]
\begin{centering}
\includegraphics[width=0.5\textwidth]{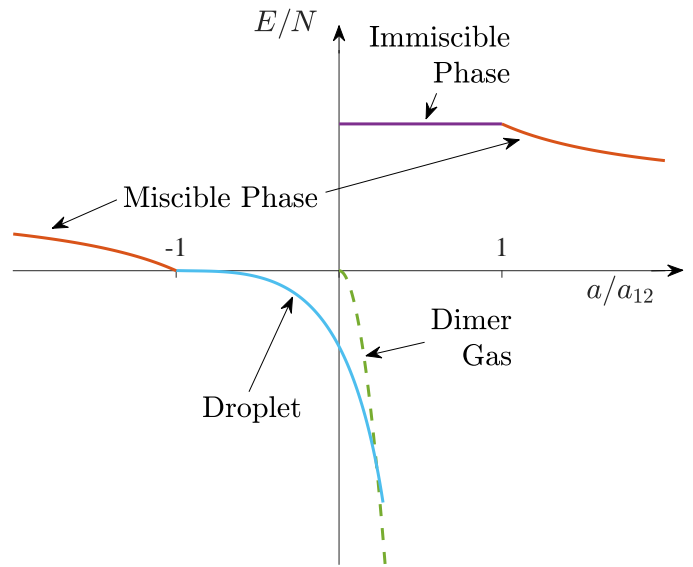} 
\par\end{centering}
\caption{\label{fig_phasediagram3d} A sketch of the phase diagram of a binary
Bose mixture across the bosonic BEC-BCS crossover, illustrated by
the energy per particle $E/N$ at a positive and small intraspecies
scattering length $0<na^{3}\ll1$.}
\end{figure}

The above observation leads us to sketch a general phase diagram for
a binary Bose mixture across the Feshbach resonance for the attractive
interspecies interactions, as illustrated in Fig. \ref{fig_phasediagram3d}.
When we increase the interspecies attraction from $a_{12}=0^{-}$
(i.e., the far-left part of the figure), the mixture is initially
in the miscible phase. Upon reaching the mean-field collapsing point
at $a_{12}=-a$, it turns into a weakly interacting Bose droplet as
experimentally observed \citep{Cabrera2018,Semeghini2018}. Across
the Feshbach resonance ($a/a_{12}=0$), the Bose droplet becomes strongly
interacting with a significant portion of Cooper pairs, whose size
is comparable to the mean-distance between bosons. By further increasing
the interspecies attraction to the threshold $a_{12,\textrm{crit}}\simeq3.6a$,
the Bose droplet disappears and changes into a gas of dimers via a
first-order transition.

It is worth noting that, since the intraspecies scattering length
$a$ is small, the interspecies scattering length $a_{12}$ at the
threshold would also be small and positive. This means that the gas
of dimers would be extremely difficult to reach adiabatically in the
time scale of actual experiments, due to its deep energy level. Instead,
we would observe a metastable state of the mixture, which is an immiscible
phase (i.e., a phase-separation phase where the two components do
not overlap in real space) at the threshold $a_{12,\textrm{crit}}\simeq3.6a$.
By further increasing the interspecies attraction towards a vanishing
scattering length ($a_{12}\rightarrow0^{+}$), the mixture will again
enter a miscible phase at $a_{12}=a$ and connect smoothly to the
miscible phase at $a_{12}=0^{-}$.

\section{One-dimensional quantum droplets}

We now turn to discuss low-dimensional quantum droplets, starting
from the one-dimensional case. In one dimension, the contact interaction
is well defined and does not need regularization. The interaction
strength can be characterized by using the dimensionless interaction
parameter, such as $\gamma=mg/(n\hbar^{2})=-2/(na)$ and $\eta=mg_{12}/(n\hbar^{2})=-2/(na_{12})$,
where 
\begin{eqnarray}
a & = & -\frac{2\hbar^{2}}{mg}<0\\
a_{12} & = & -\frac{2\hbar^{2}}{mg_{12}}>0
\end{eqnarray}
are the one-dimensional $s$-wave scattering lengths. Following Ref.
\citep{Parisi2019}, we choose the binding energy of a dimer of two
bosons in different species, i.e., 
\begin{equation}
\varepsilon_{B}=\frac{\hbar^{2}}{ma_{12}^{2}},
\end{equation}
as the units of energy, and the inverse scattering length $\left|a\right|^{-1}$
as the units of density.

\begin{figure}[t]
\begin{centering}
\includegraphics[width=0.5\textwidth]{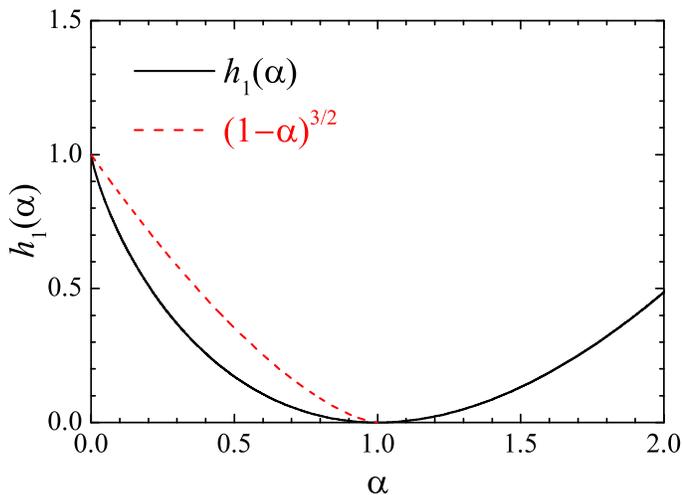} 
\par\end{centering}
\caption{\label{fig_h1A} The function $h_{1}(\alpha)$ (solid line) and its
comparison to $(1-\alpha)^{3/2}$ (dashed line).}
\end{figure}

In the Bogoliubov theory, the LHY thermodynamic potential Eq. (\ref{eq:OmegaLHYBog})
in one dimension is easy to calculate. By performing the one-dimensional
integration, 
\begin{equation}
\sum_{\mathbf{k}}\left[\sqrt{\varepsilon_{\mathbf{k}}\left(\varepsilon_{\mathbf{k}}+\alpha\right)}-\varepsilon_{\mathbf{k}}-\frac{\alpha}{2}\right]=-\frac{\left(2m\right)^{1/2}\alpha^{3/2}}{3\pi\hbar},\label{eq:indentity1d}
\end{equation}
we obtain 
\begin{equation}
\varOmega_{\textrm{LHY}}=-\frac{2m^{1/2}}{3\pi\hbar}\left[\left(C+D\right)^{3/2}+\left(C-D\right)^{3/2}\right].
\end{equation}
By taking $C=gn/2$ and $D=g_{12}n/2$ as before in $\varOmega_{\textrm{LHY}}$
and adding the mean-field energy $(g+g_{12})n^{2}/4$, we obtain the
energy per particle predicted by the Bogoliubov theory, 
\begin{equation}
\frac{E_{\textrm{1D}}}{N}=\left(g+g_{12}\right)\frac{n}{4}-\frac{\left(2m\right)^{1/2}}{6\pi\hbar}g^{3/2}\mathcal{F}_{1}\left(\frac{g_{12}}{g}\right)n^{1/2},\label{eq:EnergyBog1d}
\end{equation}
where $\mathcal{F}_{1}(\alpha)\equiv(1+\alpha)^{3/2}+(1-\alpha)^{3/2}$.
It is interesting to note that, in one dimension the LHY energy functional
is negative so the force provided by quantum fluctuations is attractive.
It is to be balanced by the repulsive mean-field force at $g>-g_{12}>0$.
Therefore, somehow counterintuitively, the formation of quantum droplets
is driven by quantum fluctuations \citep{Petrov2016}. As the mean-field
solution is stable, the energy in Eq. (\ref{eq:EnergyBog1d}) does
not suffer from the issue of complex number as we encounter earlier
in three dimensions. Nevertheless, following Ref. \citep{Petrov2016}
we may still use the Petrov's prescription and take $g_{12}=-g$ in
Eq. (\ref{eq:EnergyBog1d}) to define an energy per particle, 
\begin{equation}
\frac{E_{\textrm{1D}}}{N}=\left(g+g_{12}\right)\frac{n}{4}-\frac{2m^{1/2}}{3\pi\hbar}g^{3/2}n^{1/2}.\label{eq:EnergyPetrov1d}
\end{equation}

\begin{figure}[t]
\begin{centering}
\includegraphics[width=0.5\textwidth]{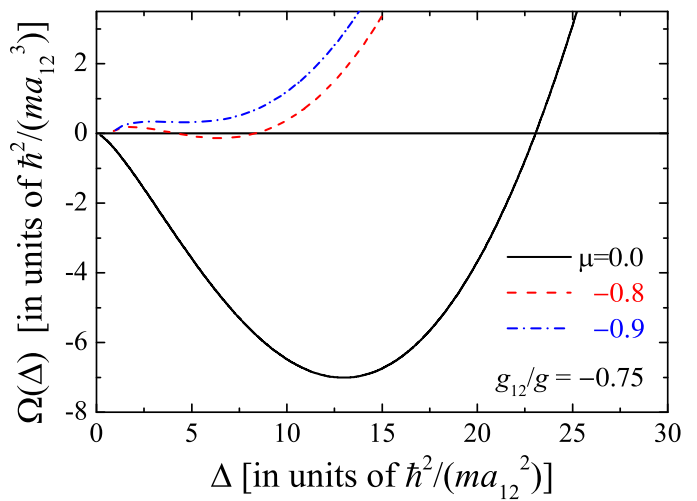} 
\par\end{centering}
\caption{\label{fig_omega1d} One-dimensional thermodynamic potential $\varOmega_{\textrm{1D}}$,
in units of $\hbar^{2}/(ma_{12}^{3})=\varepsilon_{B}a_{12}^{-1}$,
as a function of the pairing parameter $\Delta$, at different chemical
potentials $\mu=0$ (solid line), $-0.8$ (dashed line), and $-0.9$
(dot-dashed line), and at $g_{12}=-0.75g$. Both $\Delta$ and $\mu$
are measured in units of $\varepsilon_{B}$.}
\end{figure}

In the pairing theory, we calculate the thermodynamic potential Eq.
(\ref{eq:OmegaPairing}) in one dimension. As in the three-dimensional
case, we similarly separate the LHY thermodynamic potential into two
parts, $\mathcal{I}_{-}$ and $\mathcal{I}_{+}$. It is straightforward
to obtain $\mathcal{I}_{-}$ with the help of the identity Eq. (\ref{eq:indentity1d})
and we obtain, 
\begin{equation}
\mathcal{I}_{-}=-\frac{4m^{1/2}}{3\pi\hbar}C^{3/2}\left(1+\frac{\Delta}{C}\right)^{3/2}.
\end{equation}
For $\mathcal{I}_{+}$, we instead find

\begin{equation}
\mathcal{I}_{+}=-\frac{4m^{1/2}}{3\pi\hbar}C^{3/2}h_{1}\left(\alpha\right),
\end{equation}
where the function $h_{1}(\alpha)$ is given by, 
\begin{equation}
h_{1}\equiv3\int_{0}^{\infty}dt\left[t^{2}+\frac{1+\alpha}{2}-\sqrt{\left(t^{2}+1\right)\left(t^{2}+\alpha\right)}\right],
\end{equation}
and is plotted in Fig. \ref{fig_h1A}. Therefore, we obtain the total
thermodynamic potential ($C=\mu+\Delta$), 
\begin{equation}
\varOmega_{\textrm{1D}}=-\frac{C^{2}}{g}-\frac{\Delta^{2}}{g_{12}}-\frac{2m^{1/2}}{3\pi\hbar}C^{3/2}\mathcal{G}_{1}\left(\frac{\Delta}{C}\right),
\end{equation}
where $\mathcal{G}_{1}(\alpha)\equiv(1+\alpha)^{3/2}+h_{1}(\alpha)$.

In Fig. \ref{fig_omega1d}, we show the thermodynamic potential $\varOmega_{\textrm{1D}}$
as a function of $\Delta$ at the interspecies interaction strength
$g_{12}=-0.75g$. The curves at three different chemical potentials
$\mu=0$, $-0.8$, and $-0.9$, measured in units of $\varepsilon_{B}$,
are plotted. When the chemical potential is above a critical value,
i.e., $\mu_{c}\simeq-0.8\varepsilon_{B}$, we typically find a global
minimum in the thermodynamic potential at $\Delta_{0}\neq0$. For
$\mu<\mu_{c}$, the global minimum turns into a \emph{local} minimum
and hence the saddle-point pairing parameter takes the trivial solution
$\Delta_{0}=0$. As a result, there is a jump in $\Delta_{0}$ when
we tune the chemical potential across $\mu_{c}$. Physically, this
indicates a \emph{first-order} quantum phase transition from a droplet
phase to a gas-like expanding phase. In other words, when the density
$n$ is very dilute (at $\mu<\mu_{c}$), the attractive force provided
by quantum fluctuations (i.e., the LHY energy $\propto n^{3/2}$)
is unstable to balance the repulsive mean-field force (i.e., characterized
by the mean-field energy $\propto n^{2}$). Thus, the expansion of
the gas-like phase can no longer be prevented.

\begin{figure}[t]
\begin{centering}
\includegraphics[width=0.5\textwidth]{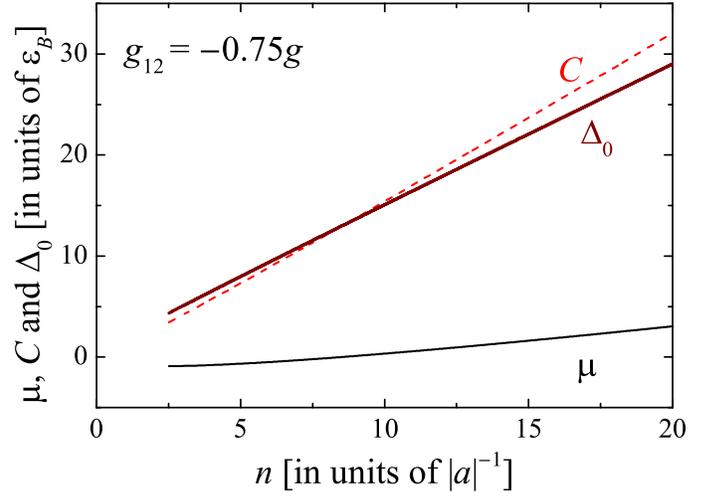} 
\par\end{centering}
\caption{\label{fig_parameters1d} One-dimensional chemical potential $\mu$,
the parameter $C$ and the pairing gap $\Delta_{0}$, in units of
$\varepsilon_{B}$, as a function of the total density $n$ (in units
of $\left|a\right|^{-1}$) at $g_{12}=-0.75g$.}
\end{figure}

By finding the saddle-point solution $\Delta=\Delta_{0}\neq0$ through
the minimization of $\varOmega(\Delta)$, we consequently calculate
the density $n=-\partial\varOmega(\mu,\Delta_{0})/\partial\mu$. The
resulting parameter $C=\mu+\Delta_{0}$ and the pairing gap $\Delta_{0}$,
together with the chemical potential $\mu$, are shown in Fig. \ref{fig_parameters1d}
as a function of the density $n$, at a typical interspecies interaction
strength $g_{12}=-0.75g$. Here, we are always in the weak-coupling
regime, since the dimensionless interaction parameters such as $\gamma=2/(n\left|a\right|)<1$.
Unlike the three-dimensional case, the condition $\left|\mu\right|\ll C,\Delta_{0}$
seems to be less satisfied at low densities, where we see a clear
difference between $C$ and $\Delta_{0}$. Therefore, although an
approximate analytical energy equation can still be derived, i.e.,
\begin{equation}
\frac{E_{\textrm{1D}}}{N}=-\left(g+\frac{g^{2}}{g_{12}}\right)\frac{n}{4}-\frac{2m^{1/2}}{3\pi\hbar}g^{3/2}n^{1/2},\label{eq:Energy1dPairingAnalytic}
\end{equation}
we would prefer to use the full numerical calculation. A comparison
between the numerical and analytical results of our pairing theory
at the interspecies interaction strength $g_{12}=-0.75g$ is shown
in Appendix A.

\begin{figure}[t]
\begin{centering}
\includegraphics[width=0.5\textwidth]{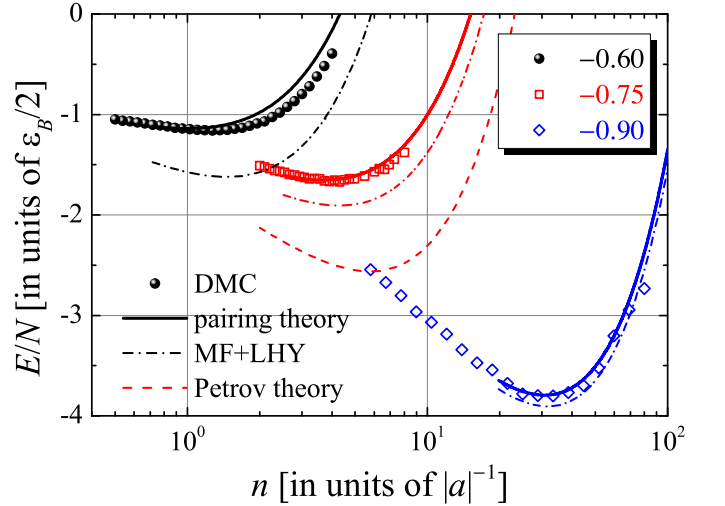} 
\par\end{centering}
\caption{\label{fig_energy1d} One-dimensional energy per particle as a function
of the density at three interspecies interaction strengths $g_{12}/g=-0.60$
(black), $-0.75$ (red) and $-0.90$ (blue). Our pairing results (solid
lines) are compared with the recent DMC data (symbols) \citep{Parisi2019},
the MF+LHY predictions Eq. (\ref{eq:EnergyBog1d}) (dot-dashed lines),
and the MF+LHY results with Petrov's prescription Eq. (\ref{eq:EnergyPetrov1d})
(dashed line, for $g_{12}/g=-0.75$ only). The energy is in units
of $\varepsilon_{B}/2$ and the density is in units of $\left|a\right|^{-1}$.}
\end{figure}

\begin{figure}[t]
\begin{centering}
\includegraphics[width=0.5\textwidth]{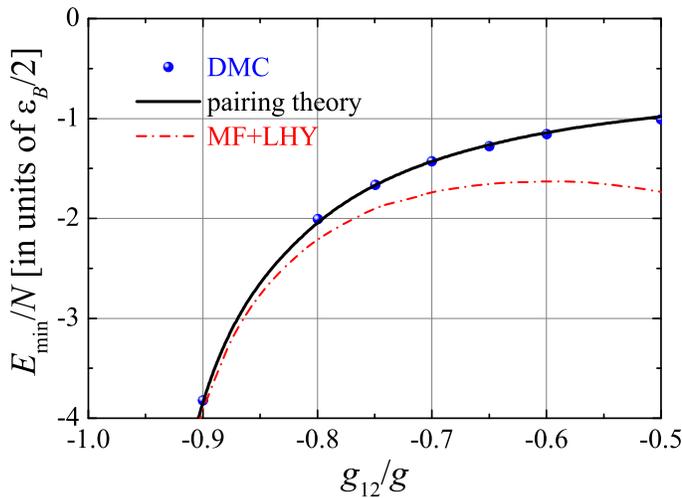} 
\par\end{centering}
\caption{\label{fig_emin1d} Minimum energy per particle $E_{\min}/N$ or equilibrium
chemical potential $\mu_{\textrm{eq}}$ in one dimension as a function
of the interspecies interaction strength $g_{12}/g$, obtained from
the pairing theory (solid line), the DMC simulation (circles) \citep{Parisi2019}
and the MF+LHY theory Eq. (\ref{eq:EnergyBog1d}) (dot-dashed line).
The energy is in units of $\varepsilon_{B}/2$.}
\end{figure}

In Fig. \ref{fig_energy1d}, we present the energy per particle predicted
by the pairing theory at three interspecies interactions $g_{12}/g=-0.60$
(black), $-0.75$ (red) and $-0.90$ (blue) by solid lines, and compare
them to the available DMC data taken from Ref. \citep{Parisi2019}
(symbols), to the Bogoliubov results Eq. (\ref{eq:EnergyBog1d}) (dot-dashed
lines), and to the Bogoliubov prediction with Petrov's prescription
Eq. (\ref{eq:EnergyPetrov1d}) (dashed line) \citep{Petrov2016}.
We find an excellent agreement between our pairing theory and the
state-of-the-art DMC simulation at all three interaction strengths.
The agreement at $g_{12}=-0.60g$ is particularly impressive, as the
dimensionless density $n\left|a\right|$ decreases and becomes close
to unity so the dimensionless interaction parameter $\gamma=2/(n\left|a\right|)\sim O(1)$
is large. We would rather anticipate the breakdown of the Bogoliubov
approximation, which our pairing theory relies on. Indeed, at this
interaction strength the conventional Bogoliubov prediction Eq. (\ref{eq:EnergyBog1d})
already shows significant deviation from the DMC data. We attribute
the good agreement between our theory and the DMC simulation to our
reasonable description of the bosonic pairing.

To understand this, it is useful show the chemical potential $\mu_{\textrm{eq}}$
at the equilibrium density (or the minimum energy per particle $E_{\min}/N$)
as a function of the interspecies interaction strength $g_{12}/g$,
as reported in Fig. \ref{fig_emin1d}. The excellent agreement between
our pairing theory and the DMC simulation for the equilibrium chemical
potential $\mu_{\textrm{eq}}$ is fairly evident, up to a critical
interspecies interaction strength $(g_{12}/g)_{\textrm{crit}}\sim-0.47(2)$
as predicted by the DMC \citep{Parisi2019}. Towards the critical
interaction strength, the equilibrium chemical potential quickly approaches
the half of the binding energy of a dimer, i.e., $-\varepsilon_{B}/2$,
indicating that the system could be understood as a collection of
weakly-interacting dimers. This interpretation is reasonable, as the
DMC threshold $(g_{12}/g)_{\textrm{crit}}\sim-0.47(2)$ is close to
the zero-crossing in the effective dimer-dimer interaction $(g_{12}/g)_{0}\sim-0.45$
\citep{Pricoupenko2018}.

Our pairing theory precisely provides a useful description of those
weakly-interacting dimers at the mean-field level, since we take a
uniform pairing gap in the saddle point solution. Thus, we anticipate
the pairing theory may predict a similar critical interspecies interaction
strength as in the DMC simulation. By determining the equilibrium
density $n_{\textrm{eq}}$ at different interspecies interactions
near the zero-crossing of dimer-dimer scattering, we find the equilibrium
density vanishes at $(g_{12}/g)_{\textrm{crit}}\sim-0.35$, which
seems to be consistent with the DMC prediction \citep{Parisi2019}
and the few-body zero-crossing result \citep{Pricoupenko2018}.

\section{Two-dimensional quantum droplets}

We finally consider two-dimensional quantum droplets. In two dimensions,
the regularization of the bare interaction strength becomes subtle
due to the logarithmic \emph{infrared} divergence at low energy, which
we may remove by introducing an \emph{arbitrary} energy scale $\varepsilon_{c}$,
i.e.,

\begin{eqnarray}
\frac{1}{g} & = & \frac{m}{4\pi\hbar^{2}}\ln\left(\frac{4\hbar^{2}}{e^{2\gamma}ma^{2}\varepsilon_{c}}\right)-\sum_{\mathbf{k}}\frac{1}{2\varepsilon_{\mathbf{k}}+\varepsilon_{c}},\label{eq:Reg2DA}\\
\frac{1}{g_{12}} & = & \frac{m}{4\pi\hbar^{2}}\ln\left(\frac{4\hbar^{2}}{e^{2\gamma}ma_{12}^{2}\varepsilon_{c}}\right)-\sum_{\mathbf{k}}\frac{1}{2\varepsilon_{\mathbf{k}}+\varepsilon_{c}}.\label{eq:Reg2DB}
\end{eqnarray}
Here, $\gamma\simeq0.577216$ is Euler--Mascheroni constant, $a$
and $a_{12}$ are two-dimensional $s$-wave scattering lengths. Alternatively,
we may consider the use of the binding energies $E_{T}\equiv4\hbar^{2}/(e^{2\gamma}ma^{2})$
and $E_{S}\equiv4\hbar^{2}/(e^{2\gamma}ma_{12}^{2})$, where analogous
to the fermionic case the subscripts ``$T$'' and ``$S$'' emphasize
the tendency of the formation of triplet and singlet pairs for bosons
in the same-species and unlike-species, respectively. We then rewrite
the bare interaction strengths in a simpler form, 
\begin{eqnarray}
\frac{1}{g} & = & -\sum_{\mathbf{k}}\frac{1}{\hbar^{2}\mathbf{k}^{2}/m+E_{T}},\\
\frac{1}{g_{12}} & = & -\sum_{\mathbf{k}}\frac{1}{\hbar^{2}\mathbf{k}^{2}/m+E_{S}}.
\end{eqnarray}
In this section, we use $E_{T}$ and $a^{-2}$ as the units of energy
and density, respectively.

In the Bogoliubov theory, the approximate energy of quantum droplets
was derived by Petrov and Astrakharchik in Ref. \citep{Petrov2016}.
For $\ln(a_{12}/a)\gg1$, it takes the form \citep{Petrov2016}, 
\begin{equation}
\frac{E_{\textrm{2D}}}{N}=\frac{2\pi n}{\ln^{2}\left(a_{12}/a\right)}\left[\ln\left(\frac{n}{n_{\textrm{eq}}}\right)-1\right],\label{eq:EnergyPetrov2d}
\end{equation}
where the equilibrium density is 
\begin{equation}
n_{\textrm{eq}}a^{2}=\frac{e^{-2\gamma-3/2}}{\pi}\frac{\ln\left(a_{12}/a\right)}{\left(a_{12}/a\right)}.
\end{equation}

\begin{figure}[t]
\begin{centering}
\includegraphics[width=0.5\textwidth]{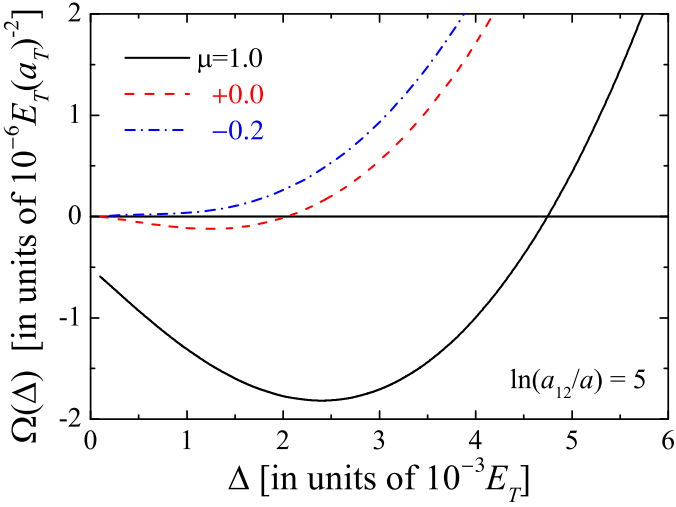} 
\par\end{centering}
\caption{\label{fig_omega2d} Two-dimensional thermodynamic potential $\varOmega_{\textrm{2D}}$,
in units of $10^{-6}E_{T}a_{T}^{-2}$, as a function of the pairing
parameter $\Delta$, at different chemical potentials $\mu=1.0$ (solid
line), $0$ (dashed line), and $-0.2$ (dot-dashed line), and at the
interspecies interaction strength $\ln(a_{12}/a)=5$. Both $\Delta$
and $\mu$ are measured in units of $10^{-3}E_{T}$.}
\end{figure}

\begin{figure}[t]
\begin{centering}
\includegraphics[width=0.5\textwidth]{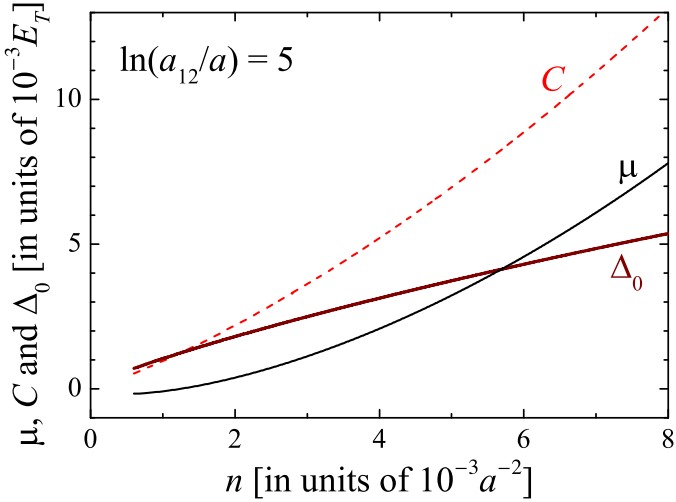} 
\par\end{centering}
\caption{\label{fig_parameters2d} Two-dimensional chemical potential $\mu$,
the parameter $C$ and the pairing gap $\Delta_{0}$, in units of
$10^{-3}E_{T}$, as a function of the total density $n$ (in units
of $10^{-3}a^{-2}$) at $\ln(a_{12}/a)=5$.}
\end{figure}

In our pairing theory, by replacing the bare interaction strengths
with the binding energies, the thermodynamic potential can be written
as ($C=\mu+\Delta$), 
\begin{eqnarray}
\varOmega_{\textrm{2D}} & = & \frac{1}{2}\sum_{\mathbf{k}}\left[E_{+}\left(\mathbf{k}\right)+E_{-}\left(\mathbf{k}\right)-2\left(\varepsilon_{\mathbf{k}}+C+\Delta\right)\right.\nonumber \\
 &  & \left.+\frac{2C^{2}}{2\varepsilon_{\mathbf{k}}+E_{T}}+\frac{2\Delta^{2}}{2\varepsilon_{\mathbf{k}}+E_{S}}\right].\label{eq:OmegaPairing2d}
\end{eqnarray}
It is interesting to see that the condensate term now disappears after
regularization. This also happens if we choose the regularization
through Eq. (\ref{eq:Reg2DA}) and Eq. (\ref{eq:Reg2DB}), since the
cut-off energy $\varepsilon_{c}$ can be arbitrarily selected. The
same trick was used in Ref. \citep{Petrov2016} to derive the Bogoliubov
result Eq. (\ref{eq:EnergyPetrov2d}). A vanishing condensate term
is related to the fact that in two dimensions, the small interaction
parameter is given by $1/\ln(na^{2})$ and one has to include the
LHY term in the energy, in order to have a meaningful perturbative
expansion expression for the energy \citep{Schick1971}. As the controlling
parameter is only logarithmically small, as we shall see, it appears
more challenging to obtain an accurate result within perturbation
theories.

The integration in Eq. (\ref{eq:OmegaPairing2d}) can be performed
analytically, as in usual two-dimensional mean-field theories. We
find that, 
\begin{eqnarray}
\varOmega_{\textrm{2D}} & = & \frac{m}{4\pi\hbar^{2}}\left[C\Delta-C_{2}\sqrt{C\Delta}+\frac{\mu^{2}}{4}\ln\left(\sqrt{C}+\sqrt{\Delta}\right)\right.\nonumber \\
 &  & \left.+\frac{C_{2}^{2}}{2}\ln\left(eC_{2}\right)-C^{2}\ln E_{T}-\Delta^{2}\ln E_{S}\right],
\end{eqnarray}
where $C_{2}\equiv C+\Delta=\mu+2\Delta$. In Fig. \ref{fig_omega2d},
we examine the $\Delta$-dependence of the thermodynamic potential
at the interspecies interaction strength $\ln(a_{12}/a)=5$. It clearly
shows a global minimum when the chemical potential is above a threshold,
similar to the three-dimensional and one-dimensional cases. Therefore,
we determine the saddle-point pairing gap $\Delta_{0}$ and consequently
calculate the density and total energy. The resulting parameter $C$
and the pairing gap $\Delta_{0}$ are shown in Fig. \ref{fig_parameters2d},
as a function of the density $n$. The chemical potential $\mu$ is
also shown. We find that with increasing the density, the chemical
potential $\mu$ increases rapidly and is larger than the pairing
gap $\Delta_{0}$ at sufficiently large densities.

\begin{figure}[t]
\begin{centering}
\includegraphics[width=0.5\textwidth]{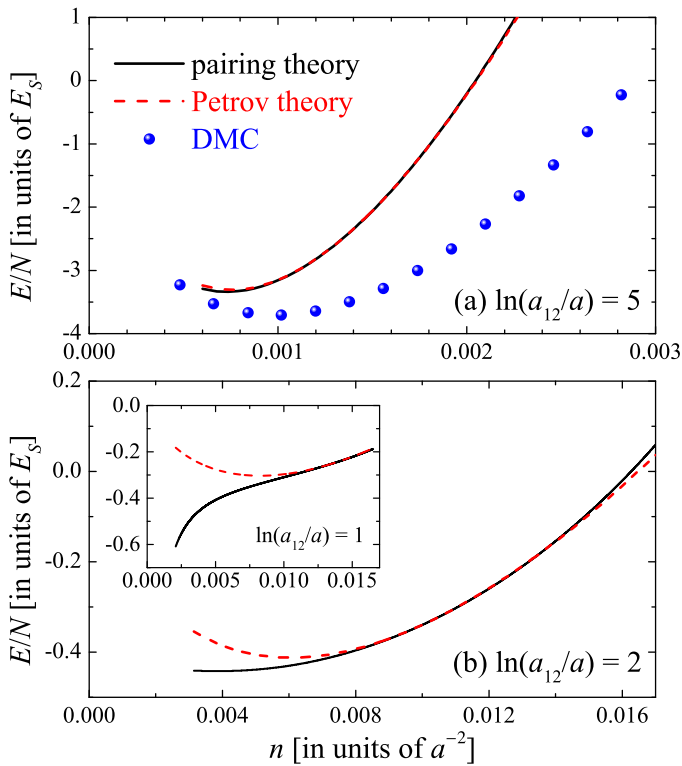} 
\par\end{centering}
\caption{\label{fig_energy2d} Two-dimensional energy per particle as a function
of the density at three interspecies interaction strengths $\ln(a_{12}/a)=5$
(a), $\ln(a_{12}/a)=2$ (b), and $\ln(a_{12}/a)=1$ (inset in (b)).
Our pairing results (solid lines) are compared with the MF+LHY predictions
with Petrov's theory Eq. (\ref{eq:EnergyPetrov2d}) (dashed line).
In (a), we show also the DMC data from Ref. \citep{Petrov2016}. The
energy is in units of the interspecies binding energy $E_{S}$ and
the density is in units of $a^{-2}$. We note that the scales of energy
and density change by a factor of $\sim10$ in the upper and lower
panels.}
\end{figure}

In Fig. \ref{fig_energy2d}, we report the density dependence of the
energy per particle of two-dimensional quantum droplets at two interspecies
interaction strengths, $\ln(a_{12}/a)=5$ (a) and $\ln(a_{12}/a)=2$
(b). Our pairing results are compared with Petrov's prediction Eq.
(\ref{eq:EnergyPetrov2d}) and the DMC data (for $\ln(a_{12}/a)=5$
only) \citep{Petrov2016}. For a weak interspecies interaction, as
shown in Fig. \ref{fig_energy2d}(a), there is a very close agreement
between our pairing result and Petrov's result. Both of them seems
to strongly over-estimate the energy, in comparison to the benchmark
DMC data, in spite of the weak interspecies interaction. This is understandable:
as we mentioned earlier, it is difficult to have accurate perturbative
expansion in two dimensions due to the logarithmically small controlling
parameter. To improve the accuracy of theoretical prediction, we need
to go beyond the Bogoliubov approximation and to obtain the correction
beyond LHY following, for example, the procedure by Mora and Castin
in Ref. \citep{Mora2009} for a scalar two-dimensional weakly-interacting
Bose gas. This will be considered in our future works.

At a larger interspecies interaction, as illustrated in Fig. \ref{fig_energy2d}(b),
the difference between the pairing result and Petrov's result becomes
noticeable. In particular, at small densities the pairing theory predicts
a lower energy. In this regime, we anticipate that the pairing effect
start to become significant, so the explicit inclusion of the bosonic
pairing, just as we consider in the pairing theory, improve the energy.

Remarkably, by further increasing the interspecies interaction, as
can be seen from the inset of Fig. \ref{fig_energy2d}(b), we find
that the energy per particle predicted by the pairing theory decreases
\emph{monotonically} with decreasing density. There is no minimum
in the energy per particle, to support a self-bound liquid-like droplet
with zero pressure in vacuum. This is not a surprising result, as
we already find the similar situation in one dimension, where the
one-dimensional quantum droplet can disappear and turn into a bright
soliton, when the interspecies attraction stronger than the threshold
$(g_{12})_{\textrm{crit}}=-g$ \citep{Tylutki2020}. By plotting the
energy curve at different interspecies interactions, we determine
a threshold in two dimensions, $[\ln(a_{12}/a)]_{\textrm{crit}}\sim1.9$,
below which the droplet changes its fundamental characters and presumably
turns into a soliton-like many-particle bound state \citep{Hammer2004,Bazak2018}.
Incidentally, this threshold is close to the zero-crossing of the
effective dimer-dimer interaction in two dimensions, i.e., $[\ln(a_{12}/a)]_{\textrm{0}}\simeq\ln(10)\simeq2.3$,
obtained from the few-body calculations \citep{Guijarro2020}.

\begin{table}[t]
\begin{centering}
\begin{tabular}{|c|c|c|}
\hline 
Dimensions  & Formation threshold  & Disappearance threshold\tabularnewline
\hline 
\hline 
One  & $\left(a_{12}/a\right)^{-1}\simeq-0.35$  & $a_{12}/a=-1$\tabularnewline
\hline 
Two  & $\ln^{-1}\left(a_{12}/a\right)=0$  & $\ln^{-1}\left(a_{12}/a\right)\simeq0.52$\tabularnewline
\hline 
Three  & $a_{12}=-a$  & $a_{12}\simeq3.6a$\tabularnewline
\hline 
\end{tabular}
\par\end{centering}
\caption{\label{TableI} Thresholds for quantum droplet formation and disappearance
in one, two and three dimensions in terms of the $s$-wave scattering
lengths, predicted by the pairing theory.}
\end{table}

\section{Conclusions}

In summary, we have presented a systematic investigation of bulk properties
of utradilute quantum droplets in a Bose-Bose mixture, by using the
recently developed pairing theory \citep{Hu2020a}. We have focused
on the low-dimensional droplets, and have found that the bosonic pairing
plays an increasingly important role in low dimensions, particularly
near the threshold at which the self-bound droplets start to emerge
or disappear, as listed in Table \ref{TableI}. We have also considered
a strongly interacting quantum droplet in three dimensions.

In one dimension, we have shown that the energy per particle predicted
by our pairing theory agrees excellent well with the numerically accurate
diffusion Monte Carlo data \citep{Parisi2019}, at all the interaction
strengths where the simulation data are available (which also nearly
cover the phase window where one-dimensional quantum droplets exist).
Our pairing theory also predicts a critical interspecies attraction
for the emergence of droplets, i.e., $(g_{12}/g)_{\textrm{crit}}\sim-0.35$,
which is consistent with the DMC prediction $(g_{12}/g)_{\textrm{crit}}\sim-0.47(2)$
\citep{Parisi2019} and with the zero-crossing point $(g_{12}/g)_{0}\sim-0.45$
where the effective dimer-dimer interaction changes from repulsive
to attractive \citep{Pricoupenko2018}.

In two dimensions, quantum droplets form for an arbitrary small interspecies
attraction. We have found our pairing theory becomes less efficient,
due to the weak interspecies attraction for pairing and the logarithmically
small controlling parameters that disfavors the development of accurate
perturbation theories. Yet, our pairing theory still provides an improvement
compared with the prototype theory of two-dimensional quantum droplets
developed earlier \citep{Petrov2016}. With increasing interspecies
attractions, the pairing theory seems to become more useful. We have
predicted a threshold $[\ln(a_{12}/a)]_{\textrm{crit}}\sim1.9$, below
which the droplet may turn into a many-particle bound state predicted
earlier by Hammer and Son \citep{Hammer2004}. Interestingly, such
a threshold is close to the zero-crossing $[\ln(a_{12}/a)]_{\textrm{0}}\simeq2.3$
of the effective dimer-dimer interaction in two dimensions found through
few-body calculations \citep{Guijarro2020}.

In three dimensions, we have shown an exciting possibility of realizing
the so-called bosonic BEC-BCS crossover, by tuning the interspecies
scattering length $a_{12}$ to be infinitely large near a Feshbach
resonance. The superfluid properties of the resulting strongly interacting
quantum droplet are to be explored. We anticipate that it may have
some universal behaviors in collective dynamics and thermodynamics,
analogous to its fermionic counterpart. Across the Feshbach resonance,
we have found that the strongly interacting quantum droplet disappears
at about $a_{12}\simeq3.6a$.

In future studies, it would be interesting to use our microscopic
pairing theory to directly investigate the profile and the collective
excitations of quantum droplets, without the use of the local density
approximation or density functional theories. These fundamental properties
are important for characterizing ultradilute quantum droplets in ultracold
atomic laboratories. 
\begin{acknowledgments}
We are grateful to Tao Shi and Zhichao Guo for stimulating discussions,
to Luca Parisi, Grigory E. Astrakharchik, and Stefano Giorgini for
sharing their DMC data. This research was supported by the Australian
Research Council's (ARC) Discovery Program, Grant No. DP170104008
(H.H.), Grants No. DE180100592 and No. DP190100815 (J.W.), and Grant
No. DP180102018 (X.-J.L). 
\end{acknowledgments}

\appendix

\section{Analytic energy expression in one dimension}

\begin{figure}[t]
\begin{centering}
\includegraphics[width=0.5\textwidth]{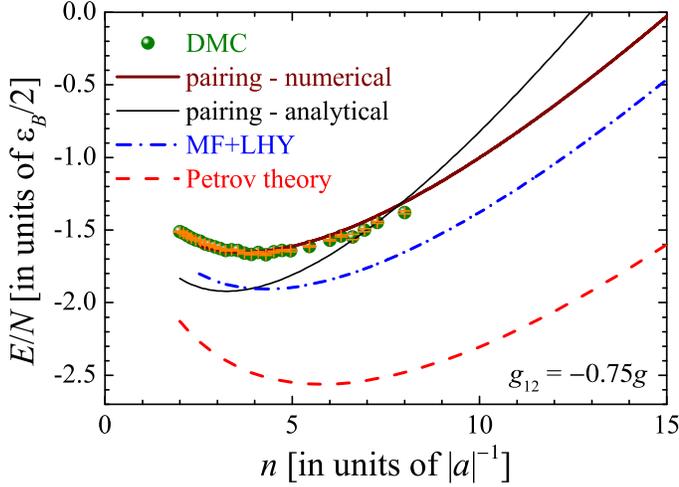} 
\par\end{centering}
\caption{\label{fig_energy1dAppendix} One-dimensional energy per particle
as a function of the density at the interspecies interaction strength
$g_{12}=-0.75g$. Our pairing results (numerical - brown thick solid
line, and analytical - black thin solid line, see Eq. (\ref{eq:Energy1dPairingAnalytic}))
are compared with the recent DMC data (symbols with error bars) \citep{Parisi2019},
the MF+LHY prediction Eq. (\ref{eq:EnergyBog1d}) (dot-dashed lines),
and the MF+LHY results with Petrov's prescription Eq. (\ref{eq:EnergyPetrov1d})
(dashed line). The energy is in units of $\varepsilon_{B}/2$ and
the density is in units of $\left|a\right|^{-1}$.}
\end{figure}

In Fig. \ref{fig_energy1dAppendix}, we show the numerical and analytical
results of our pairing theory for the one-dimensional energy per particle
at the interspecies interaction strength $g_{12}=-0.75g$. The analytical
expression Eq. (\ref{eq:Energy1dPairingAnalytic}) does not provide
a good approximation to the numerical result, since unlike in the
three-dimensional case the assumption $\left|\mu\right|\ll C,\Delta_{0}$
is not satisfied so well.

\end{document}